\documentclass[prd,12pt,nofootinbib,preprint]{revtex4}
\usepackage[T1]{fontenc}
\usepackage[utf8x]{inputenc}
\usepackage{amsmath,amssymb}
\usepackage{epsfig}
\usepackage{url}
\usepackage{subfigure}
\usepackage{graphicx}
\usepackage{grffile}
\usepackage[usenames,dvipsnames]{color}
\usepackage{slashed}
\usepackage{array}
\usepackage{multirow}
\usepackage{xcolor}
\usepackage[colorlinks,citecolor=blue]{hyperref}
\usepackage{color}
\usepackage{cancel}
\usepackage{gensymb}
\usepackage{soul}

\def \met{\not \! E_T }

\def\bar {\overline}

\def\be {\begin{equation}}
\def\ee {\end{equation}}
\def\beq {\begin{equation}}
\def\eeq {\end{equation}}
\def\bea {\begin{eqnarray}}
\def\eea {\end{eqnarray}}

\newcommand{\besub}{\begin{subequations}}
\newcommand{\eesub}{\end{subequations}}

\def\beq{\begin{equation}}
\def\eeq{\end{equation}}
\def\barr{\begin{array}}
\def\earr{\end{array}}

\def\q2 {q^2}

\def\bt{\begin{table}}
\def\et{\end{table}}

\def\mET{E_T \hspace{-1.0em}/\;\:}

\begin{document}

\title{Searches for heavy neutrinos at multi-TeV muon collider : a resonant leptogenesis perspective }

\author{Indrani Chakraborty}
\email{indranic@iitk.ac.in}
\affiliation{Department of Physics, Indian Institute of Technology Kanpur, Kanpur, Uttar Pradesh-208016, India} 

\author{Himadri Roy}
\email{himadrir@iitk.ac.in}
\affiliation{Department of Physics, Indian Institute of Technology Kanpur, Kanpur, Uttar Pradesh-208016, India}

\author{Tripurari Srivastava}
\email{tripurarisri022@gmail.com}
\affiliation{Department of Physics and Astrophysics, University of Delhi, Delhi 110007, India}

\begin{abstract} 

In this work, the standard model (SM) is extended  with two right-handed (RH) neutrinos and two singlet neutral fermions to yield active neutrino masses via (2,2) inverse see-saw mechanism. We first validate the multi-dimensional model parameter space with neutrino oscillation data, obeying the experimental bounds coming from the lepton flavor violating (LFV) decays : $\mu \to e \gamma,~ \tau \to e \gamma, ~ \tau \to \mu \gamma$. Besides we also search for the portion of the parameter space which yield the observed baryon asymmetry of the universe via resonant leptogenesis. Further, we pick up a few benchmark points from the aforementioned parameter space  with TeV scale heavy neutrinos and perform an exhaustive collider analysis of the final states : $2l + \mET$ in multi-TeV muon collider.
\end{abstract} 
\maketitle
\section{Introduction}
\label{intro}
While the Discovery of the Higgs boson at the Large Hadron Collider (LHC) \cite{Aad:2012tfa,Chatrchyan:2012ufa} completes the particle spectrum of the Standard Model (SM), this spin-zero boson also confirms the mass generation mechanism of the fermions and gauge bosons via spontaneous symmetry breaking. However, an exception of the aforesaid mechanism occurs for the neutrinos owing to the absence of the counterpart of the left-handed neutrinos in SM. On the contrary to this theoretical observation, the flavor oscillations of the neutrinos yield massive active neutrinos with an upper limit of $\mathcal{O}$(0.1 eV) \cite{Aghanim:2018eyx} coming from cosmological observations. This tiny neutrino mass can be generated via {\em see-saw mechanism}, which requires the extension of SM with additional fermionic or bosonic degrees of freedom. Depending on the nature and representations of the extended sector, various types of see-saw mechanisms like Type-I \cite{Minkowski:1977sc,Mohapatra:1979ia,Schechter:1980gr,Schechter:1981cv}, Type-II \cite{Konetschny:1977bn,Cheng:1980qt,Lazarides:1980nt,Mohapatra:1980yp,Antusch:2007km,Barbieri:1979ag,Magg:1980ut,Felipe:2013kk,Chakraborty:2019uxk,Rodejohann:2004cg,Chen:2010uc,Parida:2020sng}, Type-III \cite{Foot:1988aq,Albright:2003xb,Suematsu:2019kst,Parida:2016asc,Biswas:2019ygr}, have been studied in the literature. The requirement of tiny active neutrino masses pushes the masses of the additional beyond Standard Model (BSM) fields to higher end and also put a lower bound on the masses of the same. In most of the see-saw mechanisms, the BSM fields are too heavy to be produced and analysed in the present and future collider experiments. The inverse see-saw mechanism \cite{Mohapatra:1986aw,Mohapatra:1986bd,Bernabeu:1987gr,Gavela:2009cd,Parida:2010wq,Garayoa:2006xs,Abada:2014vea,Law:2013gma,Nguyen:2020ehj,Deppisch:2004fa,Arina:2008bb,Dev:2009aw,Malinsky:2009df,Hirsch:2009ra,Blanchet:2010kw,Dias:2012xp,Agashe:2018cuf,Gautam:2020wsd,Zhang:2021olk} turns out to be very effective for addressing this problem as it can produce TeV scale heavy neutrinos which is well within the reach of future collider experiments. In the inverse see-saw framework, SM is extended by gauge singlet right-handed neutrinos and singlet neutral fermions as will be mentioned in detail later. 

%
%
%

Another shortcoming of SM causes lack of explanation of the observed baryon asymmetry of the universe. According to the current observation \cite{Aghanim:2018eyx}, the baryon asymmetry is:
\bea
\eta_B = \frac{n_B - n_{\bar{B}}}{n_\gamma} = (6.12 \pm 0.04) \times 10^{-10} \,.
\label{baryon-limit}
\eea

Following the {\em Sakharov conditions} \cite{Sakharov:1967dj}, baryon number, $C$ and $CP$ symmetry as well as thermal equilibrium should be violated to provide an explanation to the dynamic generation of this asymmetry. One of the most popular mechanisms to generate this asymmetry is leptogenesis \cite{Fukugita:1986hr,Covi:1996wh,Roulet:1997xa,Pilaftsis:1997jf,Buchmuller:2005eh,Chun:2007vh,Kitabayashi:2007bs,Prieto:2009zz,Suematsu:2011va,AristizabalSierra:2011ab,Hambye:2012fh,Kashiwase:2013uy,Borah:2013bza,Hamada:2015xva,Zhao:2020bzx}, where the lepton asymmetry can originate from the out-of-equilibrium decay of heavy neutrinos. This asymmetry is further translated to baryon asymmetry via sphaleron transitions in SM \cite{Rubakov:1996vz,PhysRevD.30.2212,PhysRevD.28.2019}, which is basically a $B-L$ conserving but $B+L$ violating process. One of the minimal models where leptogenesis can be realised by adding heavy right-handed neutrinos (with masses $> 10^{9}$ GeV \cite{Davidson:2008bu,Davidson:2002qv}) to SM, can yield correct baryon asymmetry via Type-I thermal leptogenesis . 
Thus to probe interesting signatures involving the heavy neutrinos in various colliders, one needs to lower the masses of the heavy neutrinos atleast to the TeV scale. A specific framework called {\em inverse see-saw} (ISS) serves the aforementioned purpose, where 
two of the mass eigenstates of the heavy neutrinos are almost mass degenerate, yielding the required  baryon asymmetry via resonant leptogenesis \cite{Hambye:2001eu,Hambye_2002,Pilaftsis:2003gt,Hambye_2004,Hambye:2004jf,Pilaftsis:2005rv,Cirigliano:2006nu,Xing:2006ms}. 

In this paper, we consider a minimal inverse see-saw scenario ISS(2,2) \cite{Abada:2014vea}, where the SM is extended by two generations of right-handed neutrinos and two SM gauge singlet neutral fermions. The phenomenology of the model in light of neutrino oscillation data \cite{Esteban:2020cvm}, lepton flavor violating decay \cite{TheMEG:2016wtm} and leptogenesis, has been thoroughly studied by us in one of our previous works \cite{Chakraborty:2021azg}. Here, we intend to explore the parameter space derived from our previous study from collider perspective and examine the prospect of some interesting signatures at the multi-TeV muon collider \cite{Long:2020wfp}. As we mentioned earlier, this particular framework provides with TeV scale heavy neutrinos which can produce the baryon asymmetry of the universe in the correct ballpark, as well as can be probed at present and future colliders. Here we aim to probe the signal involving the pair production of active neutrinos along with the heavy neutrinos (masses ranging from 3.28 TeV to 10.12 TeV), followed by the subsequent decay of the heavy neutrinos to $W^\pm$ and charged leptons. Further we consider the leptonic decay of $W^\pm$, leading to $2l+ \mET$ final state. We would have been opted for LHC or ILC to search for the aforementioned signal. In the context of LHC, the signal significance turns out to be low due to the presence of huge SM background and tiny signal cross-section at 14 TeV. The significance might be improved in the proposed 100 TeV collider but still, the SM background cross-section is expected to be very large compare to the signal cross-section. Besides the similar signal involving TeV scale heavy neutrinos at the production level cannot be investigated at ILC, since the maximum achievable center of mass energy (COM) at ILC is 1 TeV. 

Now the possibility of detecting TeV scale neutrino can be achieved in a leptonic collider with a sufficiently high COM energy. Recently, there is growing interest in muon colliders with multi TeV COM energy. This can provide a clean environment with hugely achievable signal cross-section, enhancing the signal significance. For our analyses, we choose multi TeV COM energies ($\sqrt{s} = 6~{\rm TeV},14$ TeV) at the muon collider with integrated luminosity ($\mathcal{L}$) varying as \cite{Han:2021udl} : 

\bea
\mathcal{L} = \large(\frac{\sqrt{s}}{10 ~ {\rm TeV}}\large)^2 \times 10^4 ~~ {\rm fb}^{-1}.
\eea

With this motivation, we analyze the $2l + \mET$ final state at multi-TeV muon collider with $ \sqrt{s} = 6 $ TeV and $14$ TeV. We shall show how a high significance could be produced by applying suitable cuts on the relevant kinematic variables to suppress the SM background using traditional cut-based analysis.

This paper is organized as follows. In Section \ref{model}, we describe the framework and the mechanism of neutrino mass generation. Further, we study phenomenological constraints in the context of the model in Section \ref{constraints}. In Section \ref{collider}, we perform cut-based analysis to analyze the collider signature at multi-TeV muon collider. Finally, we summarize and conclude in Section~\ref{conclusion}.

\section{Model}
\label{model}
In the present work, we will be operating in the framework of SM, minimally extended with two right-handed neutrinos $N_{{R_1}}, N_{{R_2}}$ and two neutral singlet fermions $S_1, S_2$ yielding tiny neutrino mass and mixing via inverse see-saw mechanism. Table \ref{quantum_no} shows the $SU(3)_C, ~ SU(2)_L, ~ U(1)_Y$ quantum numbers assigned to the bosonic and fermionic fields of the model. Here the hyper-charge $Y$ is calculated using : $Q = T_3 + \frac{Y}{2}$, where $T_3$ and $Q$ are the weak isospin and electric charge of the field respectively. In Table \ref{quantum_no}, the SM Higgs doublet is denoted by $\phi$. $Q_{L_i}, L_{L_i}$ are the left-handed SM quark and lepton doublets respectively, whereas $u_{R_i}, d_{R_i}, \ell_{R_i}$ are right-handed up-type, down-type quark and lepton singlets respectively with $i=3$.

\begin{table}[htpb!]
\begin{center}
\begin{tabular}{|c|c|c|c|}
\hline
\hspace{5mm} Particles \hspace{5mm} &  \hspace{5mm} $SU(3)_C$ \hspace{5mm} & \hspace{5mm} $SU(2)_L$ \hspace{5mm} &  \hspace{5mm} $U(1)_Y$ ~~\hspace{5mm}\\ \hline \hline
$\phi$ & 1 & 2 & 1 \\ \hline
$Q_{L_i} = \begin{pmatrix}
u_{L_i}\\
d_{L_i}
\end{pmatrix}, ~ i=3$  & 3 & 2 & $\frac{1}{3}$ \\ \hline

$u_{R_i}, ~ i=3$ & 3 & 1 & $\frac{4}{3}$\\
\hline 
$d_{R_i}, ~ i=3$ & 3 & 1 & -$\frac{2}{3}$\\ \hline 
$L_{L_i} = \begin{pmatrix}
\nu_{L_i}\\
\ell_{L_i}
\end{pmatrix}, ~ i=3$ & 1 & 2 & -1\\
\hline 
$\ell_{R_i}, ~ i=3$ & 1 & 1 & -2\\ \hline 
$N_{R_j}, ~ j=2$ & 1 & 1 & 0 \\
\hline 
$S_j, ~ j=2$ & 1 & 1 & 0 \\ \hline
\end{tabular}
\end{center}
\caption{$SU(3)_C, ~ SU(2)_L, ~ U(1)_Y$ quantum number assigned to the particles. $i$ denotes the number of generations.}
\label{quantum_no}
\end{table}
The Yukawa Lagrangian signifying the inverse see-saw mechanism is :

\bea
-\mathcal{L}_y = y_{i \alpha} \bar{N}_{R_i} \phi^\dag \ell_{L_\alpha} + \frac{1}{2} M_{R_{ij}} N_{R_i}^T C^{-1} N_{R_j}+ M_{S_{ij}} N_{R_i}^T C^{-1} S_j + \frac{1}{2} \mu_{ij} S_i^T C^{-1} S_j + {\rm h.c.}
\label{lagrangian_flavor}
\eea
where $\alpha$ represents the flavor of leptons. Here $y_{i \alpha}$ denotes the Yukawa coupling matrix with complex entries and $C$ is the charge conjugation operator. The aforementioned Lagrangian contains both lepton number conserving Dirac mass term, as well as lepton number violating Majorana mass terms.

In the flavor basis $(\nu_L^i, N_{R_j}^c, S_k)^T$ (with $i=3,~ j,k=2$), following Eq.(\ref{lagrangian_flavor}), the neutrino mass matrix can be written as :

\bea
M_\nu =\begin{pmatrix}
0 & 0 & 0 & {M_D}_{1,1} & {M_D}_{1,2} & 0 & 0 \\
0 & 0 & 0 & {M_D}_{2,1} & {M_D}_{2,2} & 0 & 0 \\
0 & 0 & 0 & {M_D}_{3,1} & {M_D}_{3,2} & 0 & 0 \\
{M_D}_{1,1} & {M_D}_{2,1} & {M_D}_{3,1} & {M_R}_{1,1} & {M_R}_{1,2} & {M_S}_{1,1} & {M_S}_{1,2} \\
{M_D}_{1,2} & {M_D}_{2,2} & {M_D}_{3,2} & {M_R}_{1,2} & {M_R}_{2,2} & {M_S}_{2,1} & {M_S}_{2,2} \\
0 & 0 & 0 & {M_S}_{1,1} & {M_S}_{2,1} & \mu_{1,1} & \mu_{1,2} \\
0 &  0 & 0 & {M_S}_{1,2} & {M_S}_{2,2} & \mu_{1,2} & \mu_{2,2}
\end{pmatrix}
\label{Mnu}
\eea

Here each and every matrix element of $M_\nu$ is considered to be complex to generalise the analysis, and thus the elements can be decomposed into real and imaginary parts as :
\bea
&& {M_D}_{i,j} = {M_D}_{i,j}^R + i ~{M_D}_{i,j}^I ~, ~~{M_R}_{l,m} = {M_R}_{l,m}^R + i ~{M_R}_{l,m}^I ~, \nonumber \\
&& {M_S}_{a,b} = {M_S}_{a,b}^R + i ~{M_S}_{a,b}^I ~, ~~ \mu_{p,q } = \mu_{p,q}^R + i ~\mu_{p,q}^I
\eea
We can write down the neutrino mass matrix $M_\nu$ in a compact form as :

\bea
M_\nu = \begin{pmatrix}
0 & M_D & 0 \\
M_D^T & M_R & M_S \\
0 & M_S^T & \mu 
\end{pmatrix} 
\label{neutrino-mat}
\eea
with $M_D = y_{i\alpha} \frac{v}{\sqrt{2}}$. With three generations of active neutrinos, two generations of $N_{R_j}$ and $S_j$s, $M_\nu$ is $7\times7$ dimensional. The individual dimensions of $M_D, M_R, M_S$ and $\mu$ are $3 \times 2, ~ 2 \times 2, ~ 2 \times 2, ~ 2 \times 2$ respectively. 

Following the mass hierarchy $\mu, M_R << M_D << M_S$, with the see-saw approximation in the inverse see-saw mechanism, the effective neutrino mass matrix looks like : \cite{CentellesChulia:2020dfh},

\begin{eqnarray}
m_\nu &=& -\begin{pmatrix}
M_D & 0
\end{pmatrix} \begin{pmatrix}
M_R & M_S \\
M_S^T & \mu
\end{pmatrix}^{-1} \begin{pmatrix}
M_D^T \\
0
\end{pmatrix} \,. \nonumber \\
&=& -M_D ~ (M_R - M_S~ \mu^{-1} M_S^T)^{-1} ~ M_D^T \,.
\label{mass-mat}
\end{eqnarray}
Thus one can neglect $M_R$ in Eq.(\ref{mass-mat}) and approximate the active neutrino mass matrix as :

\bea
m_\nu &=&  M_D ~ (M_S^T)^{-1} ~\mu ~M_S^{-1} ~M_D^T.
\label{mneutrino}
\eea

 From Eq.(\ref{mneutrino}), it is evident that, with $M_R = 0$  \footnote{In \cite{Chakraborty:2021azg}, we have shown that following the hierarchy $\mu, M_R << M_D << M_S$ , with a non-zero but tiny $M_R$, all the numerical results hardly show any
deviation with respect to what is obtained by setting $M_R = 0$. Thus our assumption of
taking a vanishing $M_R$, for simplifying the analysis, is thus justified.}, double suppression by the mass scale $M_S$ along with small $\mu$, yield tiny active neutrino mass.

  Upon diagonalising $m_\nu$ in Eq.(\ref{mneutrino}) the light neutrino masses are generated via the transformation : 
\bea 
U_{\rm PMNS}^T~ m_\nu ~U_{\rm PMNS} = {\rm diag}(m_1,m_2,m_3) = \hat{m_\nu} \,.
\label{mass_diag}
\eea
 where $m_1, m_2, m_3$ are three light active neutrino masses, $U_{\rm PMNS}$ is the Pontecorvo-Maki-Nakagawa-Sakata matrix (PMNS matrix) \footnote{$U_{\rm PMNS}$ can be written as :
\bea
U_{\rm PMNS} = \begin{pmatrix}
c_{13}c_{12} & c_{13} s_{12} & s_{13}e^{-i \delta_{\rm CP}} \\
-s_{12}c_{23}-c_{12}s_{23}s_{13}e^{i \delta_{CP}} & c_{12} c_{23}-s_{12} s_{23} s_{13} e^{i \delta_{\rm CP}}& s_{23} c_{13} \\
s_{12} s_{23} - c_{12} c_{23} s_{13}e^{i \delta_{\rm CP}} & -c_{12} s_{23} - s_{12} c_{23} s_{13} e^{i \delta_{\rm CP}} & c_{23} c_{13}
\end{pmatrix},
\label{UPMNS}
\eea
Here $c_{ij} \equiv \cos \theta_{ij} , s_{ij} \equiv \sin \theta_{ij}$ and $\delta_{CP}$ is the $CP$-violating phase.}. 

\section{Constraints}
\label{constraints}
\subsection{Neutrino data fitting}
Following Eq.(\ref{mneutrino}), $\mu$ can be expressed in terms of the matrices $ m_\nu, M_R, M_D, M_S$ as :
\bea
\mu = M_S^T~ M_D^{-1}~ m_\nu ~(M_D^T)^{-1}~ M_S
\label{Mmu-fit}
\eea
Here $m_\nu$ can be expressed in terms of $U_{\rm PMNS}, \hat{m_\nu}$ from Eq.(\ref{mass_diag}), where the elements of the $U_{\rm PMNS}$ matrix in Eq.(\ref{UPMNS}) are already constrained from neutrino oscillation data \cite{Esteban:2020cvm}. Thus the model parameter space which we consider in the present analysis becomes compatible with the neutrino oscillation data \footnote{Assuming normal hierarchy (NH) among the light neutrinos, the parameters are fixed at their central values :
\bea
&&\sin^2 \theta_{12} = 0.304 , ~ \sin^2 \theta_{23} = 0.573 , ~ \sin^2 \theta_{13} = 0.02219, \nonumber \\
&&\Delta m_{21}^2 = 7.42 \times 10^{-5} {\rm eV}^2 , ~ \Delta m_{31}^2 = 2.517 \times 10^{-3} {\rm eV}^2 , ~ \delta_{\rm CP} = 197^\circ \,.
\label{central-value}
\eea
}.
The texture of $M_\nu$ in Eq.(\ref{neutrino-mat}) yields one massless active neutrino, which is an unavoidable feature of the (2,2) inverse see-saw realising framework. Thus we set the lightest active neutrino mass $m_1$ to be zero, satisfying $ (m_1 + m_2 + m_3) \leq 0.12$ eV \cite{Aghanim:2018eyx,Vagnozzi:2017ovm} \footnote{$m_1, m_2, m_3 $ are the masses of three active neutrinos.}.

\subsection{Lepton Flavor violation}
Through the diagonalisation of mass matrix $M_\nu$ in Eq.(\ref{neutrino-mat}) by a $7 \times 7$ unitary matrix $U$, one can obtain seven mass eigenstates $\nu_1', \nu_2', \nu_3', \tilde{\Psi_1},\tilde{\Psi_2},\tilde{\Psi_3},\tilde{\Psi_4}$ with masses $m_1,m_2,m_3,M_{\tilde{\Psi_1}},M_{\tilde{\Psi_2}},M_{\tilde{\Psi_3}},M_{\tilde{\Psi_4}}$ respectively. Following the mass hierarchy ($\mu << M_D << M_S$) in the inverse see-saw framework, the mass difference : $|M_{\tilde{\Psi}_{1(3)}}-M_{\tilde{\Psi}_{2(4)}}| \sim \mu$. Thus the corresponding pairs having such a tiny mass difference become mass degenerate.

Thus the lepton flavor violating (LFV) decays like $l_i \rightarrow \ l_j \gamma$ obtain additional contributions coming from the heavy neutrinos $\tilde{\Psi_1},\tilde{\Psi_2},\tilde{\Psi_3},\tilde{\Psi_4}$ \cite{Chakraborty:2021azg}. Among all LFV decays the most stringent bound comes from $\mu \to e \gamma$ \cite{TheMEG:2016wtm}.

\subsection{Leptogenesis and baryon asymmetry}

\begin{figure}[htpb!]{\centering
\subfigure[]{
\includegraphics[scale=0.15]{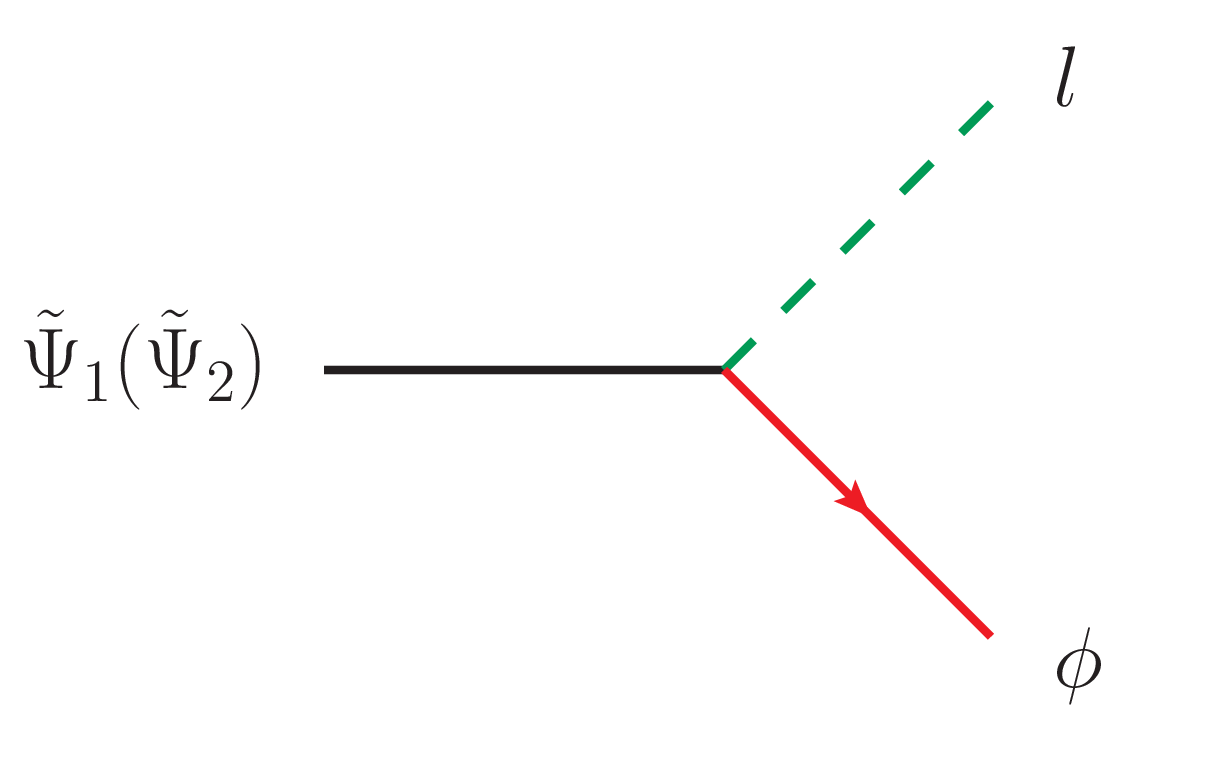}}
\subfigure[]{
\includegraphics[scale=0.15]{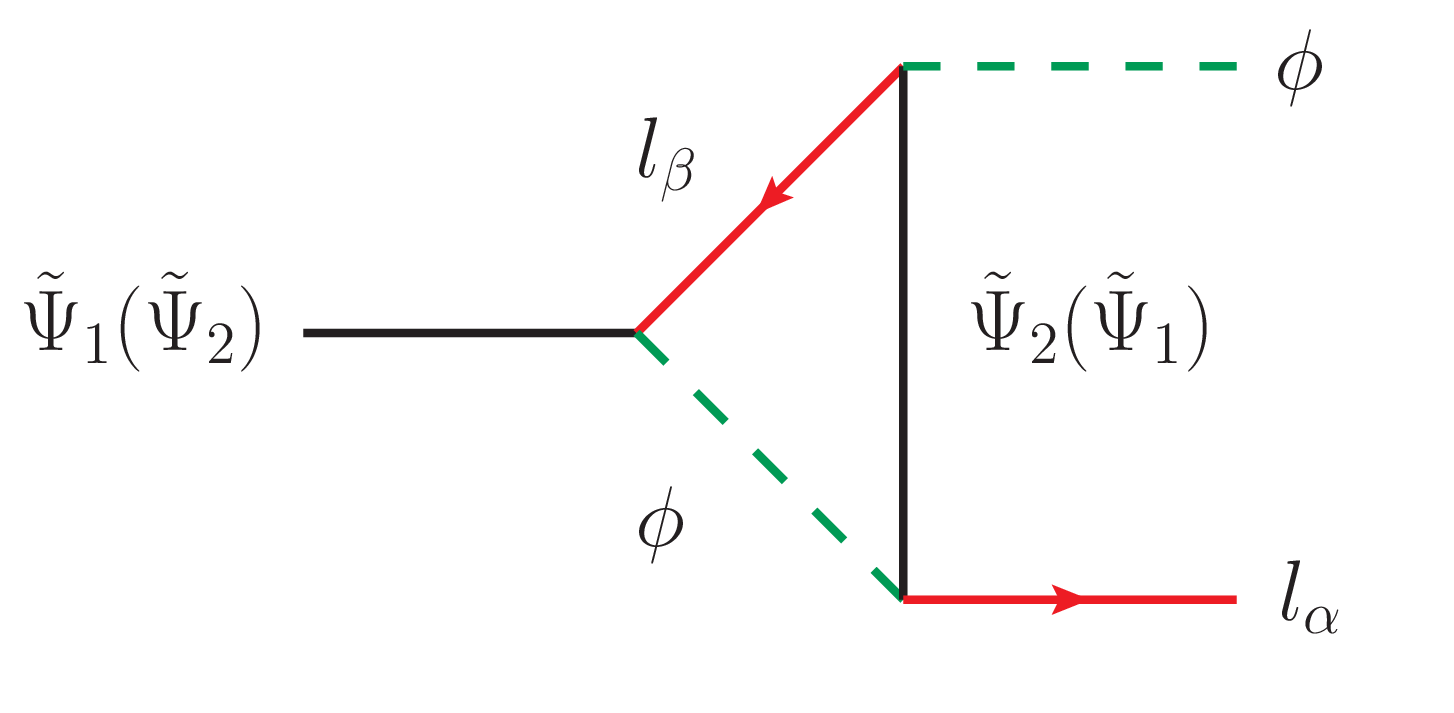}}
\subfigure[]{
\includegraphics[scale=0.15]{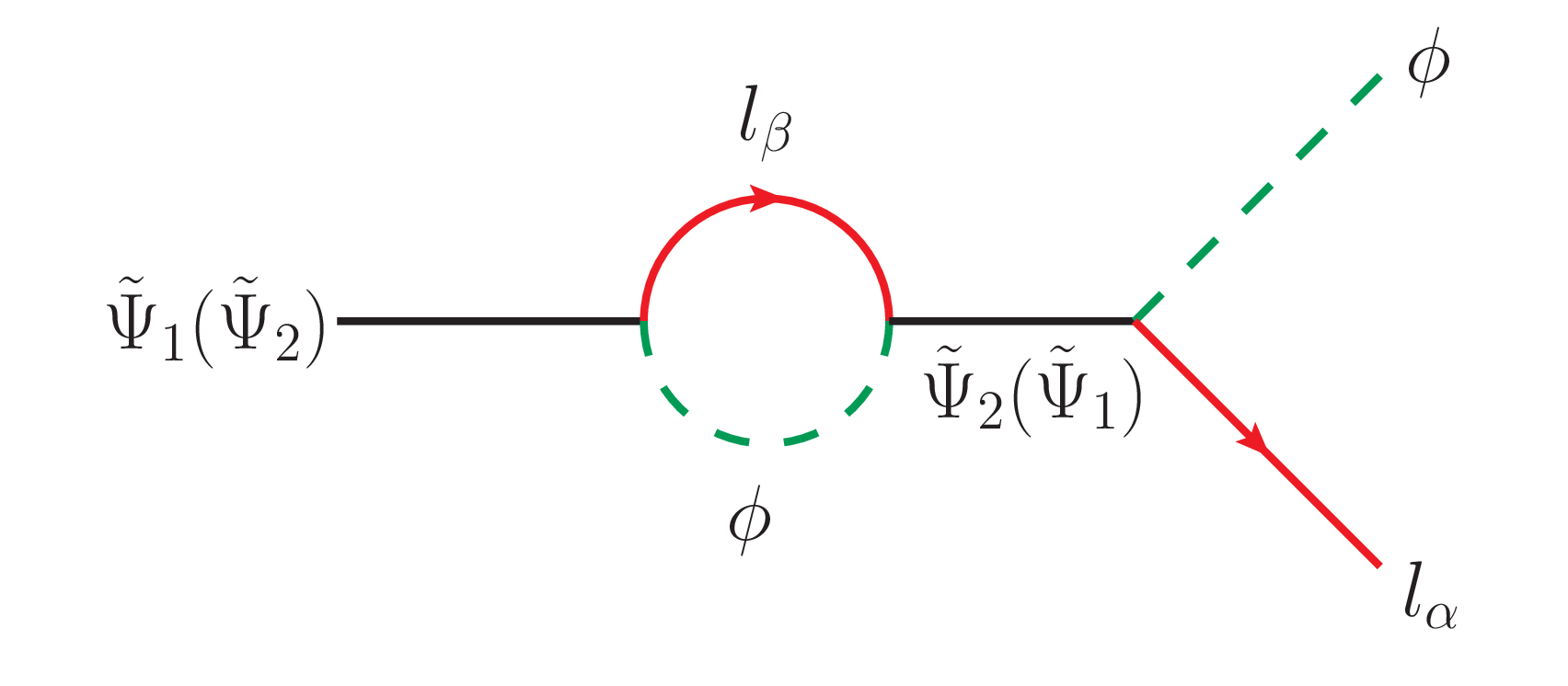}}} 
\caption{Diagrams contributing to the $CP$-asymmetry $\epsilon_1$ and $\epsilon_2$: (a) tree-level decay of $\tilde{\Psi_{1}} (\tilde{\Psi_{2}})$, (b) vertex correction, (c) self-energy diagram.}
\label{cp-asymmetry}
\end{figure}

In this subsection, we briefly discuss the generation of baryon asymmetry via leptogenesis. For an exhaustive description of the mechanism, we refer to reference \cite{Chakraborty:2021azg}. Considering the mass degeneracy of the pair of heavy neutrinos, the required CP-asymmetry will be generated through the out of equilibrium decay of the lightest mass degenerate pair ($\tilde{\Psi_{1}}, \tilde{\Psi_{2}}$) via {\em resonant leptogenesis}. The diagrammatic representation of the aforementioned processes can be found in Fig.\ref{cp-asymmetry}.

Assuming $M_R=0$, $M_\nu$ becomes block diagonalisable. While computing CP-asymmetry, the preferred choice of basis is that where the lower block of the 
$M_\nu$ (4 × 4 complex symmetric sub-matrix $\mathcal{M}$ \footnote{4 × 4 complex symmetric sub-matrix $\mathcal{M}$ is defines as :
\bea
\mathcal{M} = \begin{pmatrix}
0 & M_S \\
M_S^T & \mu
\end{pmatrix}
\eea}) is diagonal. In this preferred basis, the Lagrangian in Eq.(\ref{lagrangian_flavor}) can be rewritten as, 
\bea
-\mathcal{L}_h = h_{i \alpha} \bar{\tilde{\Psi_i}} \phi^\dag \ell_{L_\alpha} + \frac{1}{2} M_{\rm diag} \tilde{\Psi_i}^T C^{-1} \tilde{\Psi_i} + {\rm h.c.}
\label{lagrangian_mass}
\eea

The relations connecting Yukawa couplings in the diagonal mass basis ($h_{i \alpha}$) and the Yukawa couplings in the flavor basis ($y_{i \alpha}$) can be found in \cite{Chakraborty:2021azg}. The total CP-asymmetry $\epsilon_j$ in the decay of $\tilde{\Psi_j}$ into $\ell_\alpha \phi~ (\bar{\ell_\alpha} \phi^\dag)$ can be calculated by summing over the SM flavor $\beta$ as follows :

\bea
\epsilon_j = \frac{\sum_\beta \left[ \Gamma(\tilde{\Psi_j} \rightarrow \ell_\beta \phi) - \Gamma(\tilde{\Psi_j} \rightarrow \bar{\ell}_\beta \phi^\dag) \right]}{\sum_\beta \left[ \Gamma(\tilde{\Psi_j} \rightarrow \ell_\beta \phi) + \Gamma(\tilde{\Psi_j} \rightarrow \bar{\ell}_\beta \phi^\dag) \right]} = \frac{1}{8 \pi} \sum_{i\neq j} \frac{{\rm Im}[(h h^\dag)_{ji}^2]}{(h h ^\dag)_{jj}} f_{ji}
\eea

Since we are operating in the regime of resonant leptogenesis, the dominant contribution to $f_{ij}$ comes from self energy correction, {\em i.e. $f_{ij} \sim f_{ij}^{\rm self}$}, where $f_{ij}^{\rm self} = \frac{(M_i^2 - M_j^2) M_i M_j}{(M_i^2 - M_j^2)^2 + R_{ij}^2}$ \footnote{In the analysis we have considered $R_{ij} = |M_i \Gamma_i + M_j \Gamma_j|$ \cite{Chakraborty:2021azg} and $\Gamma_j = \frac{(h h^\dag)_{ii} M_j}{8 \pi}$ is the total decay width of $\tilde{\Psi_j}$. The logic behind choosing this particular form of $R_{ij}$ can be found in one of our previous studies \cite{Chakraborty:2021azg}.}.

Here we have assumed that $\tilde{\Psi_{1}}, \tilde{\Psi_{2}}$ are almost mass degenerate and lighter than other two states $\tilde{\Psi_{3}}, \tilde{\Psi_{4}}$. Thus the computed CP-asymmetry $\epsilon_1$ and $\epsilon_2$    By solving three coupled Boltzmann equations simultaneously, one can obtain co-moving densities 
$Y_{\tilde{\Psi_{1}}}, Y_{\tilde{\Psi_{2}}}$ and $Y_{B-L}$ of $\tilde{\Psi_{1}}, \tilde{\Psi_{2}}$ and $B-L$ asymmetry respectively.  The co-moving density is denoted as the ratio of actual number density \footnote{The number densities of particles with mass M and temperature T can be written as :
\bea
N_{eq} = \frac{g M^2 T}{2 \pi^2}K_2(\frac{M}{T})
\eea
$g$ being the number of degrees of freedom of corresponding particles, $K_2$ being second modified Bessel function of second kind.} and the entropy density $\bar{s}$ of the universe \footnote{Entropy density is computed as : $\bar{s} = \frac{2 \pi^2}{45}~ g_{eff} T^3$. Here $T$ is the temperature and $g_{eff}$ is the number of degrees of freedom (D.O.F), which is computed in Appendix \ref{app:A}.}. The detailed presentation and discussion of the Boltzmann equations are relegated to Appendix \ref{app:A}.

\section{Collider studies}
\label{collider}
\begin{figure}[htpb!]{\centering
\subfigure[]{
\includegraphics[width=8cm,height=6cm, angle=0]{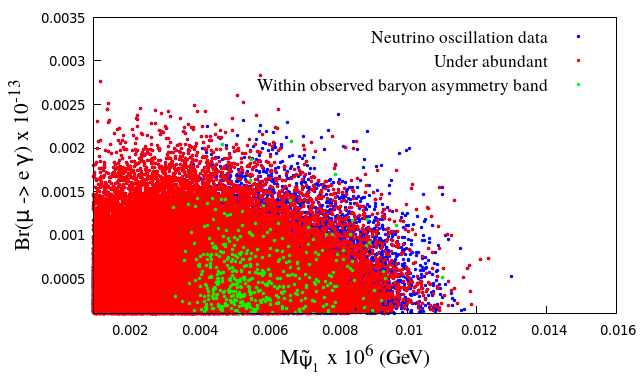}} 
\subfigure[]{
\includegraphics[width=8cm,height=6cm, angle=0]{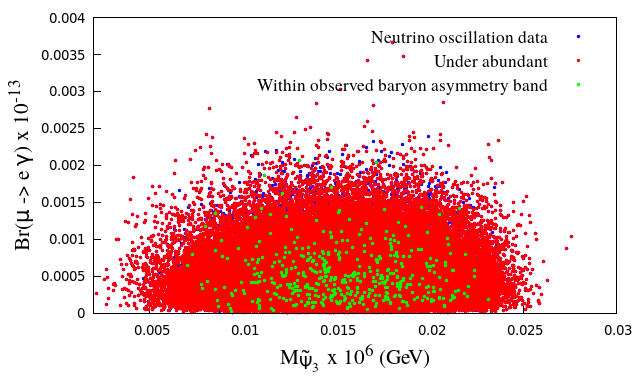}}} 
\caption{Branching ratios of radiative decay of muon ($\mu \rightarrow e \gamma$) with the mass of the heavy neutrinos ((a)$M_{\tilde{\Psi_1}}$, (b) $M_{\tilde{\Psi_3}}$). Color codes represent points satisfy the neutrino oscillation data (blue), observed baryon asymmetry (green) and under abundance (red).}
\label{Br:m1}
\end{figure}

Before proceeding to perform the collider analysis, let us discuss the nature of the model parameter space in light of various constraints described in Section \ref{constraints}. In Section \ref{model}, we already have mentioned that with the inverse see-saw hierarchy, one can ascertain tiny active neutrino masses by setting $M_R = 0$. We have also verified that this assumption hardly alters the results with respect to $M_R \neq 0$ scenario. Thus we set $M_R =0$ throughout this analysis, which in turn introduces a resemblance with the original inverse see-saw model. In addition, all the entries of $M_\nu$ (except $M_R$) are considered as complex to make the analysis a general one. While fitting the neutrino oscillation data, we adopt {\em normal hierarchy} among the light active neutrinos and also set the mass of the lightest active neutrino to be zero ($m_1 = 0$), which arises as an artifact of the present framework. For the collider analysis, we shall restrict ourselves to a particular mass region, where the mass of the lightest heavy neutrino ($M_{\tilde{\Psi_1}}$) is less than or equal to 10 TeV, {\em i.e.}  $M_{\tilde{\Psi_1}} \leq ~10$ TeV.  Now, the ranges for input parameters for the aforementioned range of $M_{\tilde{\Psi_1}} $ are : $ M_{D_{i,j}}^{R,I} \in [10^7~{\rm eV} : 10^8~{\rm eV}]$ ,  $M_{S_{i,j}}^{R,I} \in [10^{10}~{\rm eV} : 10^{14}~{\rm eV}]$. From Eq.(\ref{Mmu-fit}) one can solve $\mu$ and express it in terms of the input parameters $M_D, M_S$ ( with $M_R = 0$ ) and neutrino oscillation parameters. In addition, it has been verified that in presence of the TeV scale heavy neutrinos, the most stringent bound coming from the LFV decay $\mu \to e \gamma$ \cite{TheMEG:2016wtm}, can be easily evaded throughout the parameter space. Next by solving the Boltzmann equations (given in Appendix \ref{app:A}), one can compute the baryon asymmetry of the universe at each and every point of the parameter space.  In Fig.\ref{Br:m1}(a) (Fig.\ref{Br:m1}(b)), we have plotted Br($\mu \to e \gamma$) vs. $M_{\tilde{\Psi_1}}~(M_{\tilde{\Psi_3}}) $. The blue, red and green points are compatible with neutrino oscillation data only, neutrino oscillation data + LFV decay and neutrino oscillation data + LFV decay + current baryon asymmetry of the universe respectively. Here we have considered a 1$\sigma$ deviation around the central value of observed baryon asymmetry (Eq.(\ref{baryon-limit})).

In one of our previous studies \cite{Chakraborty:2021azg}, we have shown that there exists a stringent lower bound of $\sim$ 3.2 TeV on $M_{\tilde{\Psi_1}}$ to satisfy the current baryon asymmetry data. For performing collider analysis, we shall choose eight benchmark points (BP1, BP2, BP3, BP4, BP5, BP6, BP7, BP8) from the model parameter space, which lie well within the correct baryon asymmetry band. Thus the chosen benchmark points are compatible with the experimental data coming from neutrino oscillation, LFV decays and current baryon asymmetry of the universe. These benchmark points are characterised by low, medium and high lightest heavy neutrino masses ($M_{\tilde{\Psi_1}}$), spanning a wide range, from 3.2 TeV to 10.12 TeV. The eight benchmarks are tabulated in Table \ref{table:masses} along with  $M_{\tilde{\Psi}_{1(2)}}$, $M_{\tilde{\Psi}_{3(4)}}$ and corresponding baryon asymmetry yield.  To probe the benchmark points, we perform the collider analysis at two COMs ($\sqrt{s}$), {\em i.e.} $\sqrt{s} = 6$ TeV (for BP1, BP2, BP3) , 14 TeV (for BP4, BP5, BP6, BP7, BP8) at muon collider. 

In the next subsection we shall perform the collider analysis of the $2l + \mET$ final state, which turns out to be promising at aforementioned COMs. Owing to small signal cross section, even with high luminosity, this signal turns out to be non-promising at LHC.  We shall provide a comparative study with LHC later. Besides, this signal involving TeV scale heavy neutrinos in the final state cannot be probed at ILC since the maximum achievable COM at ILC is 1 TeV. The leading order (LO) signal and background cross sections are generated via {\texttt MG5aMC@NLO} \cite{Alwall:2014hca}. Further decays of the unstable particles are emulated through \texttt{Pythia8} \cite{Sjostrand:2014zea}. The detector effects are included in the analysis by passing both the signal and the backgrounds through Delphes-3.5.0 \cite{deFavereau:2013fsa}. We use the  default {\em muon collider} simulation card \cite{muoncollidercardTalk} for this purpose. Since traditional cut-based analysis is enough to yield high significances for all benchmarks, we shall only present the results generated from it. The signal significance has been computed using $\mathcal{S} = \sqrt{2\Big[(S + B) \log\Big(\frac{S + B}{B}\Big)- S\Big]}$ \cite{Cowan:2010js}, where $S (B)$ denote the number of signal (background) events surviving the cuts applied on relevant kinematic variables.

\begin{table}[htbp!]
	\centering
	\resizebox{10cm}{!}{
	\begin{tabular}{|c|c|c|c|c|}
		\hline
		Benchmark  & $M_{\tilde{\Psi}_{1}}  {\displaystyle \simeq }M_{\tilde{\Psi}_{2}} $  & $M_{\tilde{\Psi}_{3}} {\displaystyle \simeq }M_{\tilde{\Psi}_{4}} $  & Baryon asymmetry ($ Y_{B} $) \\ 
	Points &$(\rm TeV)$ &$(\rm TeV)$ &  \\ \hline
	
		BP1 & 3.28  & 12.31 & 8.677$\rm \times 10^{-11}$  \\ \hline
		BP2 & 4.01  & 20.18 & 8.884 $\rm \times 10^{-11}$  \\ \hline
		BP3 & 5.11  & 16.69 & 8.639 $\rm \times 10^{-11}$ \\ \hline
		BP4 & 6.33  & 16.23 & 8.708 $\rm \times 10^{-11}$ \\ \hline
		BP5 & 7.38  & 16.64 & 8.522 $\rm \times 10^{-11}$ \\ \hline
		BP6 & 8.24  & 19.23 & 8.767 $\rm \times 10^{-11}$ \\ \hline
		BP7 & 9.31 & 17.49 & 8.688 $\rm \times 10^{-11}$ \\ \hline
		BP8 & 10.12  & 17.18 & 8.713 $\rm \times 10^{-11}$\\ \hline
		
	\end{tabular}}
	\caption{Masses of Heavy Neutrinos along with the baryon asymmetry for eight benchmarks.}
	\label{table:masses}
\end{table}

\subsection{$2l + \mET$ final state}
In muon collider, heavy neutrinos can be produced along with active neutrinos via  $t$-channel $W^\pm$ mediation and $s$-channel $Z$ mediation. In this analysis, we aim to probe  the following signal shown in Fig.\ref{diagram}(a),(b) leading to $2l + \met$ {\footnote{Here, $l = e, \mu$}} final state:
\bea
&& \mu^+ \mu^- \to \nu_l ~\tilde{\Psi}_{1(2)}; \nonumber \\
&& \tilde{\Psi}_{1(2)} \to W^{\pm}~ l^{\mp}; ~  W^{\pm} \to l^{\pm}  \nu_l(\bar{\nu}_l); \nonumber \\
&& \to ~ l^+ l^- + \met
\eea
\begin{figure}[htpb!]{\centering
  \subfigure[]{
 \includegraphics[scale= 0.5, angle=0]{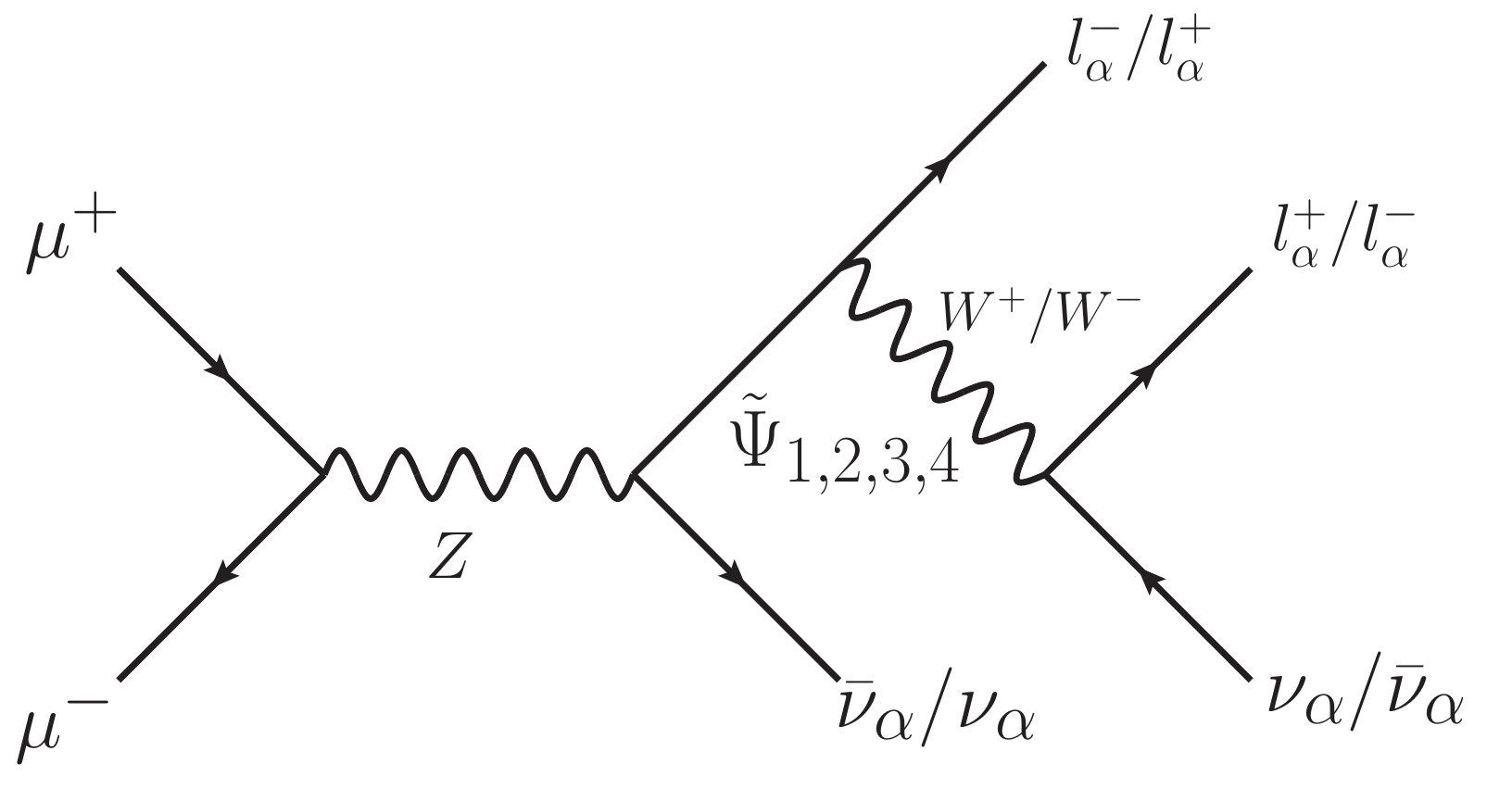}}
  \subfigure[]{
 \includegraphics[scale= 0.5, angle=0]{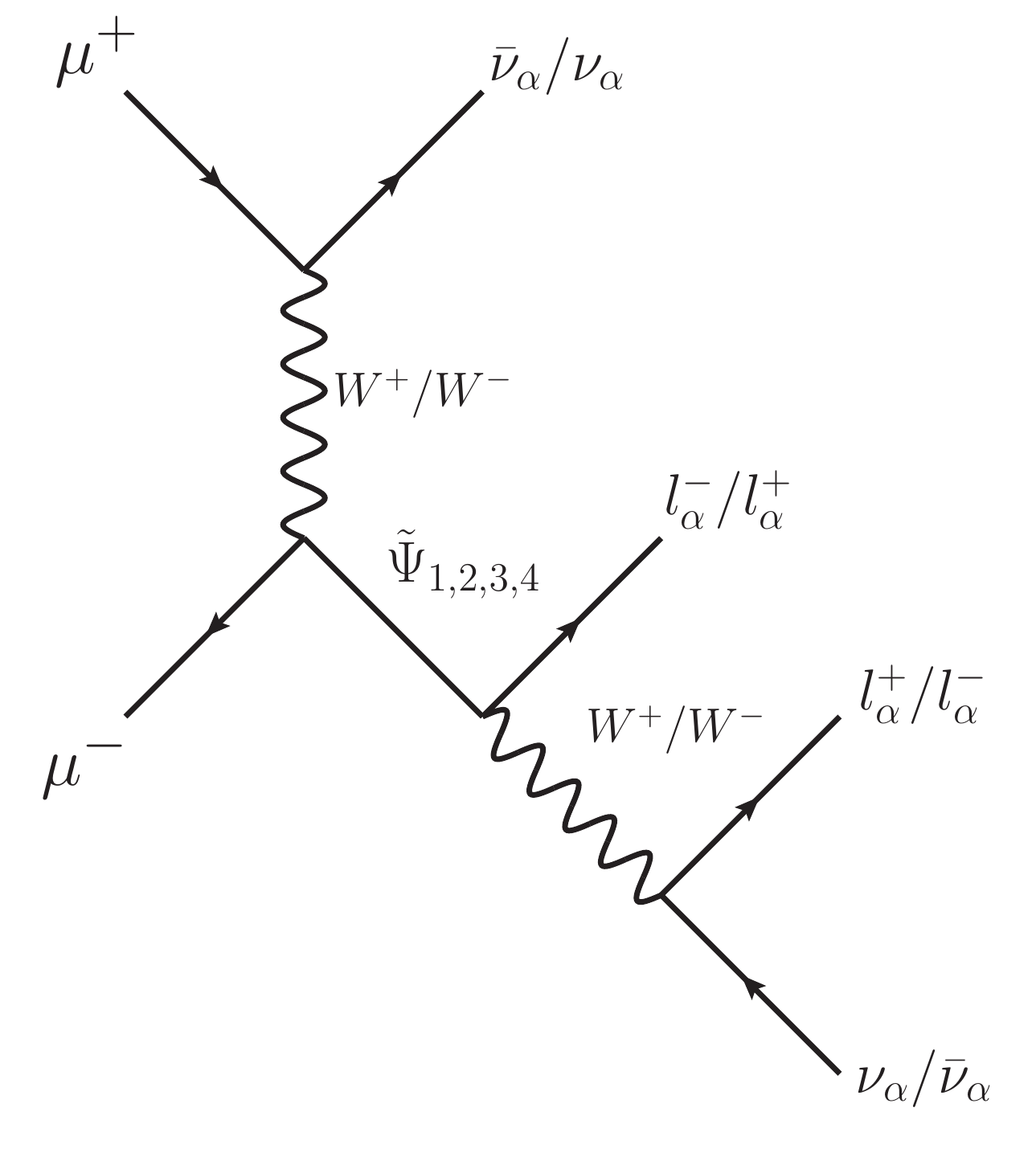}} 
  }
 \caption{The Feynman diagrams representing the signal process $ \mu^+ \mu^- \rightarrow 2l+\mET$. In these diagrams, $l^{\pm}_{\alpha}$ denote $e, \mu $. }
 \label{diagram}
 \end{figure}

Here, the heavy neutrino can decay to $W^{\pm}$ via charged current interaction which is further accompanied by the leptonic decay of $W^{\pm}$. Here we consider the leptonic decay mode of $W^{\pm}$ as it can provide a clean signature. The dominant background for this signal originates from the $\mu^+ \mu^- \to l^+ l^- + \met$ final state, which receives contributions from the following sub-processes :
\begin{itemize}
\item $\mu^+ \mu^- \to W^+ W^-$; $W^+ \to l^+ ~\nu_\ell, ~ W^-  \to l^- ~\bar{\nu}_\ell$,
\item $\mu^+ \mu^- \to ZZ$; $Z \to l^+ l^-, ~ Z  \to \nu_l ~\bar{\nu}_l$,
\item $\mu^+ \mu^- \to W^+ W^- Z$; $W^+ \to l^+ ~\nu_l, ~ W^-  \to l^- ~\bar{\nu}_l, ~ Z  \to \nu_l ~\bar{\nu}_l$,
\item $\mu^+ \mu^- \to ZZZ$; $Z \to l^+ l^-, ~ Z  \to \nu_l ~\bar{\nu}_l, ~ Z  \to \nu_l ~\bar{\nu}_l$.
\end{itemize}
\begin{table}[htbp!]
	\centering
	\resizebox{12cm}{!}{
	\begin{tabular}{|c|c|c|}
		\hline
		Signal / Backgrounds  & Process & Cross section $\sigma$ (LO) (fb)     \\ \hline
		\multicolumn{1}{|c}{ }  &\multicolumn{1}{c}{6 TeV} &  \multicolumn{1}{c|}{} \\ \cline{1-3}
		Signal & & \\ \hline
		\multicolumn{1}{|c|}{BP1 } & \multicolumn{1}{c|}{} & \multicolumn{1}{c|}{  244}    \\ 
		\multicolumn{1}{|c|}{BP2 } & \multicolumn{1}{|c|}{$\mu^{+} \mu^{-} \rightarrow 2l+\mET$} & \multicolumn{1}{|c|}{ 36}    \\ 
		\multicolumn{1}{|c|}{BP3 } & \multicolumn{1}{|c|}{} & \multicolumn{1}{|c|}{  3}    \\ \hline
		
Background & $\mu^{+} \mu^{-} \rightarrow 2l+\mET$ & 233 \\ \hline

\multicolumn{1}{|c}{ }  &\multicolumn{1}{c}{14 TeV} &  \multicolumn{1}{c|}{} \\ \cline{1-3}

		Signal & & \\ \hline
		\multicolumn{1}{|c}{BP4 } & \multicolumn{1}{|c|}{} & \multicolumn{1}{c|}{44.95}    \\ 
		\multicolumn{1}{|c}{BP5 } & \multicolumn{1}{|c|}{} & \multicolumn{1}{c|}{8.29}    \\ 
		\multicolumn{1}{|c}{BP6 } & \multicolumn{1}{|c|}{$\mu^{+} \mu^{-} \rightarrow 2l+\mET$} & \multicolumn{1}{c|}{ 2.33}    \\ 
		\multicolumn{1}{|c}{BP7 } & \multicolumn{1}{|c|}{} & \multicolumn{1}{c|}{ 0.78}    \\ 
		\multicolumn{1}{|c}{BP8 } & \multicolumn{1}{|c|}{} & \multicolumn{1}{c|}{ 0.18}    \\ \hline
 		Background & $\mu^{+} \mu^{-} \rightarrow 2l+\mET$ & 187.34 \\ \hline
	\end{tabular}}
	\caption{The LO cross sections for signal and backgrounds for the process $\mu^{+} \mu^{-} \rightarrow 2l+\mET$ at 6 TeV and 14 TeV. }
	\label{table:cross-sec}
\end{table}

\begin{figure}[htpb!]{\centering
  \subfigure[]{
 \includegraphics[scale= 0.28, angle=0]{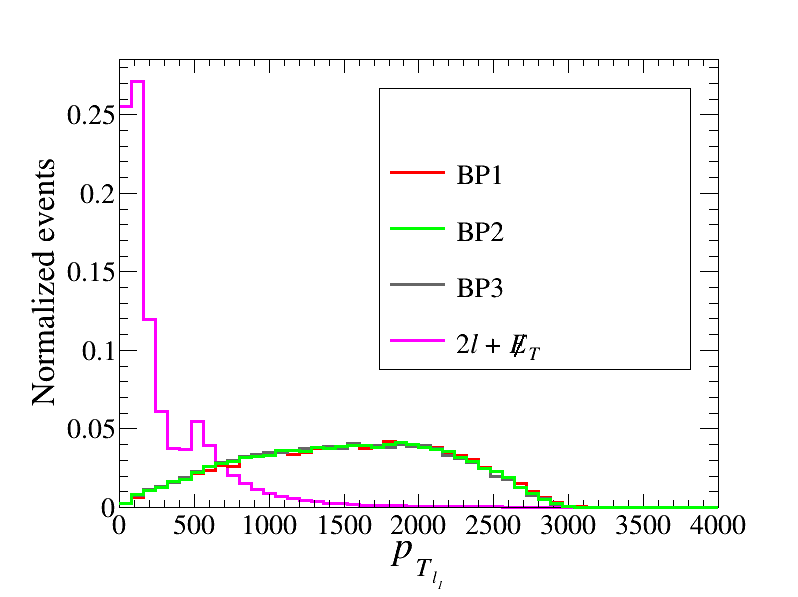}}
  \subfigure[]{
 \includegraphics[scale= 0.28, angle=0]{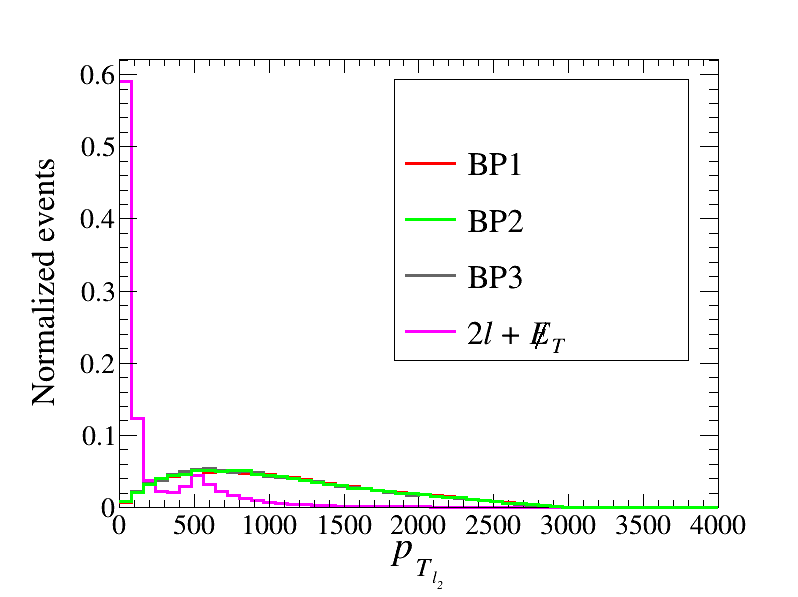}} \\
 \subfigure[]{
 \includegraphics[scale= 0.28, angle=0]{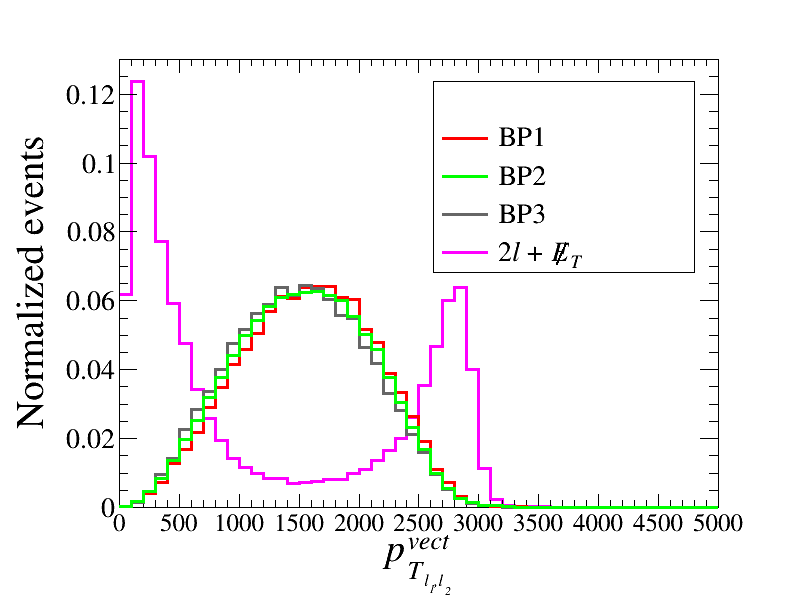}}
   \subfigure[]{
 \includegraphics[scale= 0.28, angle=0]{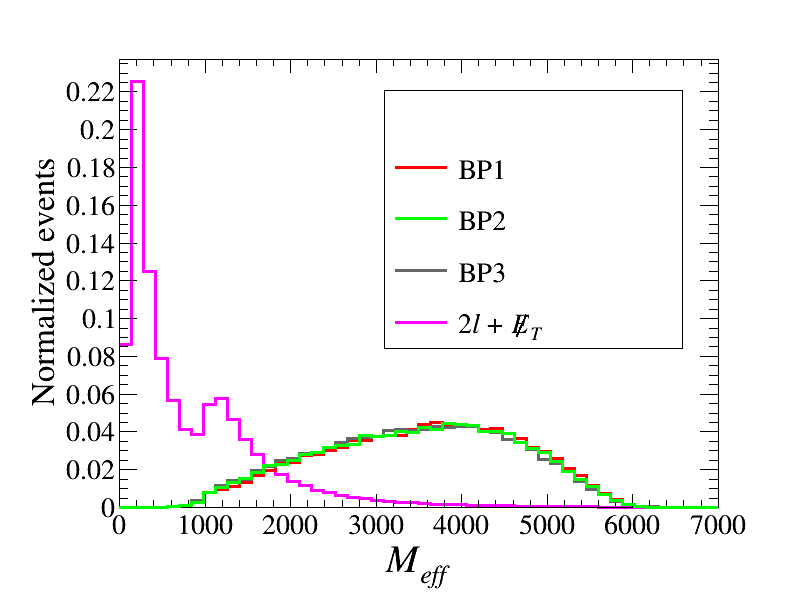}}\\
  \subfigure[]{
 \includegraphics[scale= 0.28, angle=0]{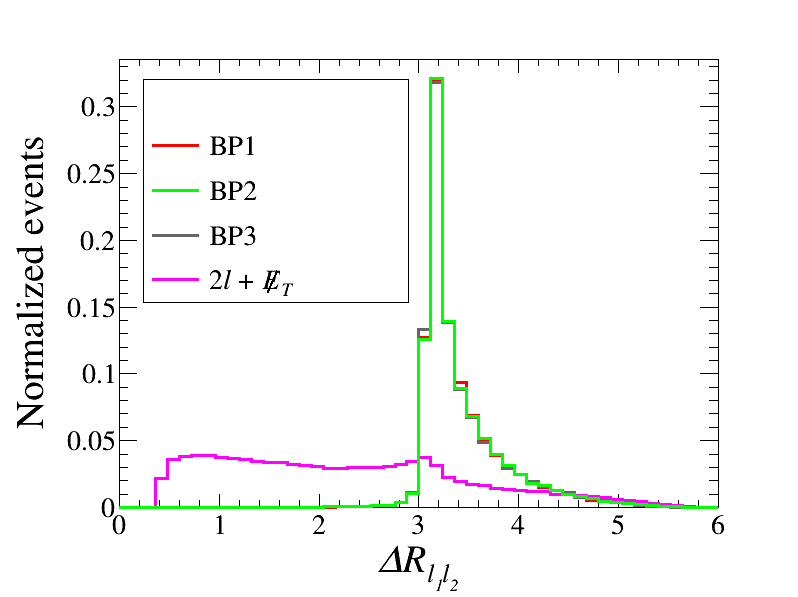}} 
  \subfigure[]{
 \includegraphics[scale= 0.28, angle=0]{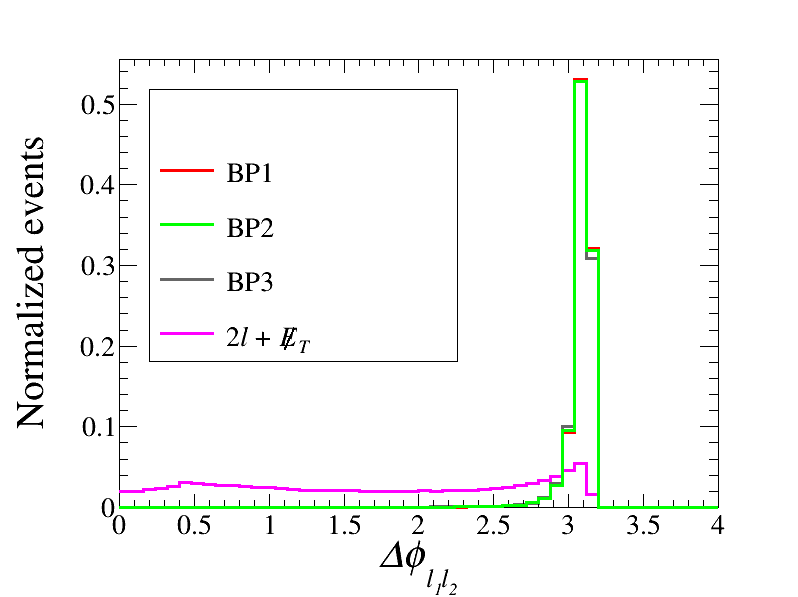}} 
  }
 \caption{ Distributions of normalized events for signal and SM background considering first three benchmarks (BP1-BP3) at 6 TeV COM, with the kinematic variables: (a) $p_{T_{l_1}}$, (b) $p_{T_{l_2}}$, (c) $p^{vect}_{T_{l_1,l_2}}$, (d) $M_{\text{eff}}$, (e) $\Delta R_{l_1 l_2}$, (f) $\Delta \phi_{l_1 l_2}$. }
 \label{fig1}
 \end{figure}

\begin{figure}[htpb!]{\centering 
   \subfigure[]{
 \includegraphics[scale= 0.28, angle=0]{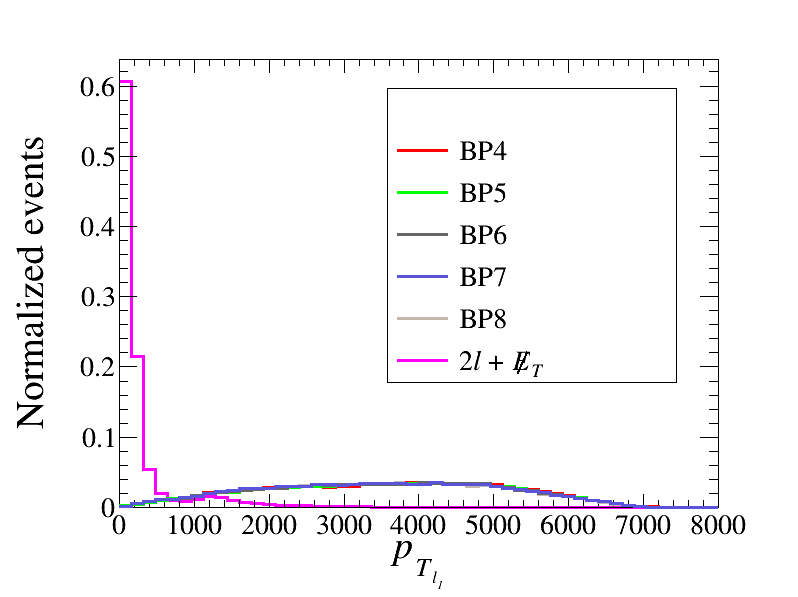}}
  \subfigure[]{
 \includegraphics[scale= 0.28, angle=0]{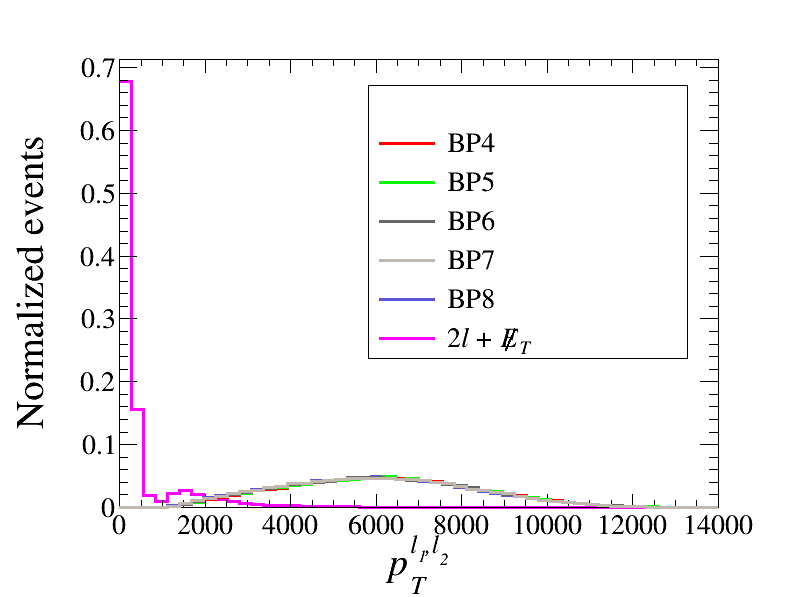}} \\ 
 \subfigure[]{
 \includegraphics[scale= 0.28, angle=0]{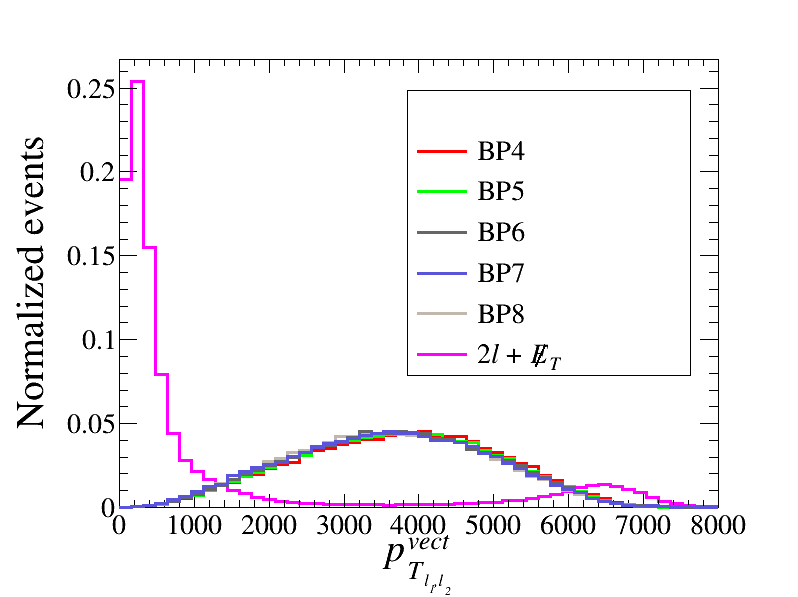}} 
 \subfigure[]{
 \includegraphics[scale= 0.28, angle=0]{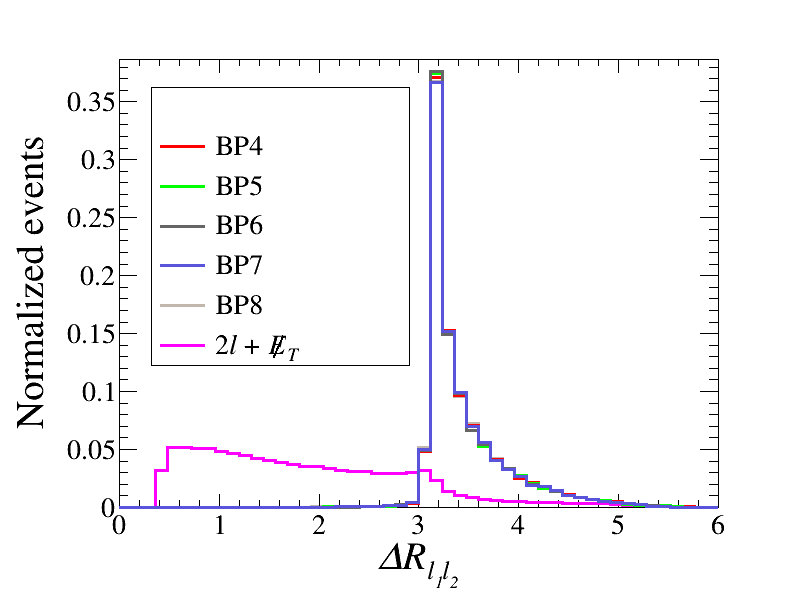}}\\
  \subfigure[]{
 \includegraphics[scale= 0.28, angle=0]{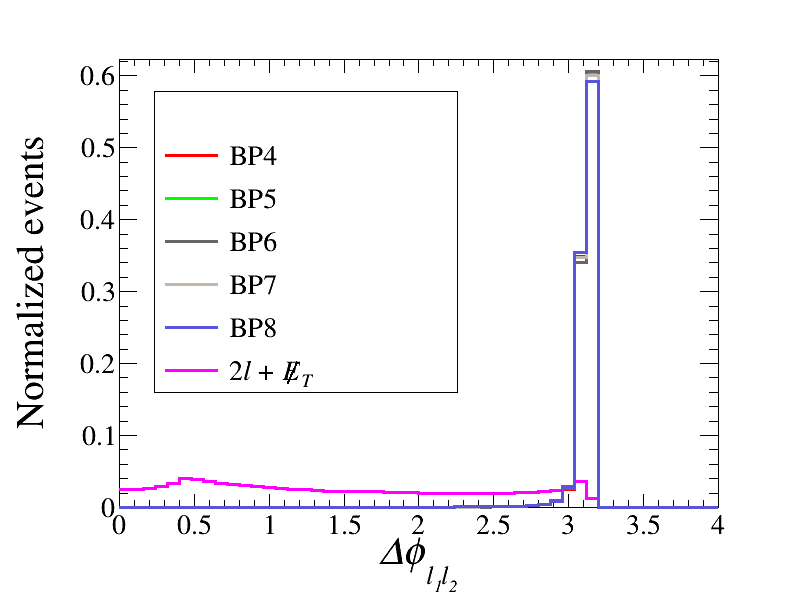}} 
   \subfigure[]{
 \includegraphics[scale= 0.28, angle=0]{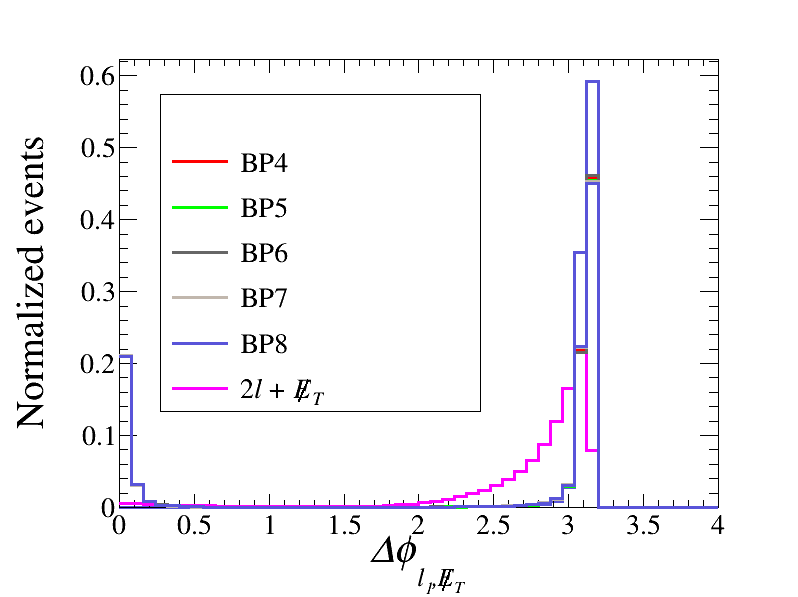}}
  }
 \caption{ Distributions of normalized events for signal and SM background considering benchmarks (BP4-BP8) at 14 TeV COM, with the kinematic variables: (a) $p_{T_{l_1}}$, (b) $p_T^{l_1,l_2}$, (c) $p_{T_{l_1 l_2}}^{vect}$, (d) $\Delta R_{l_1 l_2}$, (e) $\Delta \phi_{l_1 l_2}$, (f)  $\Delta \phi_{l_1 \cancel{E}_T}$}. 
 \label{fig4}
 \end{figure}

The LO cross sections of the signals (for all benchmarks) and SM backgrounds are tabulated in Table \ref{table:cross-sec}. To reduce the background with respect to signal, we choose a set of cuts over kinematic variables. We classify the cuts on the kinematic variables for benchmarks BP1- BP3 at 6 TeV COM energy as $B_i~(i:1-7)$ and for the rest of the benchmarks (BP4-BP8) at $\sqrt{s}=14$ TeV, the cuts are defined as $C_j~(j:1-7)$. For brevity, and also to avoid repetition, we discuss the cuts for two different COM energies together. All the names of the cuts or numbers within the bracket below will correspond to 14 TeV COM energy. Following are the depiction of cuts chosen to enhance significance:

\begin{itemize}
\item ${B_1~(C_1):}$ We demand two opposite sign, same or different flavored charged leptons $l$ in the final state. Here $l$ denotes only electron and/or muon.

\item ${B_2~(C_2):}$ The normalized distributions of the transverse momentum of the leading lepton ($p_{T_{l_1}}$) for signal and background are shown in Fig.\ref{fig1}(a) (Fig.\ref{fig4}(a)). Here the leading lepton gets the boost in $p_T$ from the COM energy directly for the signal. Thus $p_{T_{l_1}}$ for the signal peaks at a larger value with respect to the SM background. 
Thus putting a large lower cut on the transverse momentum, {\em i.e.} $p_{T_{l_1}}> 400~(200)$ GeV can be a suitable choice to enhance the signal significance.

\item $B_3$: We show the normalized distributions of the transverse momentum of the next to leading lepton ($p_{T_{l_2}}$) for signal and backgrounds in Fig.\ref{fig1}(b). As can be seen from the distributions, the SM background peaks around $p_T\sim 0$, because the major fraction of the transverse momentum is taken away by the leading lepton. The signal distribution starts dominating the background distribution after 200 GeV. Therefore, we put a lower cut of $p_{T_{l_2}}>200$ GeV to maximise the significance.

\item $ C_3:$ We depict the  normalized distributions of scalar sum of the transverse momenta of two leptons in the final state in Fig.\ref{fig4}(b). The nature of the distributions of the signal and background can be explained following the descriptions of the distributions of $p_{T_{l_1}}$ and $p_{T_{l_2}} $ in previously defined cuts B2(C2) and B3. A suitable lower cut : $p_T^{l_1,l_2}>2$ TeV helps magnifying the significance drastically. 
     
\item $B_4~(C_4):$ In Fig.\ref{fig1}(c) (Fig.\ref{fig4}(c)), we plot the normalised distributions of the magnitude of the vector sum of the transverse momenta of two leptons in the final state for signal and background. It is defined as:
     \begin{equation*}
        p^{vect}_{T_{l_1,l_2}} \equiv\sqrt{(p_{{T x}_{l_1}}+p_{{T x}_{l_2}})^2+(p_{{T y}_{l_1}}+p_{{T y}_{l_2}})^2+(p_{{T z}_{l_1}}+p_{{T z}_{l_2}})^2} 
     \end{equation*}

Here, $p_{{T x}_{l_i}}, p_{{T y}_{l_i}}$ and $p_{{T z}_{l_i}}$  are the $x,y,z$-components of the transverse momentum vector for $i=1,2$ respectively. 
We choose $p^{vect}_{T_{l_1,l_2}}>600(2000)$ GeV to suppress SM  background.

\item $B_5:$ We define a kinematic variable $M_{eff}$, {\em i.e.} effective mass as the scalar sum of the transverse momenta of the final state leptons and missing transverse energy. The normalized distributions of $M_{eff}$ both for signal and background are shown in Fig.\ref{fig1}(d). The distributions mimic the distributions for individual transverse momentum of the final state leptons as expected. Since the signal distributions for almost all benchmarks start overshadowing the SM background distribution for $M_{eff}>2000$ GeV,  to achieve large significance,  we choose {$M_{eff}>2000$ GeV}.

\item $B_6~(C_5)$:
We portray the normalized distribution of $\Delta R_{l_1 l_2}$ \footnote{ The separation between the final state leptons can be defined as:
    \[\Delta R_{l_1 l_2}=\sqrt{(\eta_{l_1}-\eta_{l_2})^2+(\phi_{l_1}-\phi_{l_2})^2},\] where $\eta_{l_i}$  and $\phi_i$ are the pseudo rapidity and azimuthal angle of $i$-th lepton.} for signals and backgrounds in Fig.\ref{fig1}(e)(Fig.\ref{fig4}(d)). It is evident from the distributions of the signal and background, a proper cut of $\Delta R_{l_1 l_2}>3.0(3.0)$ has been chosen to distinguish the signal from the background.

\item $B_7~(C_6)$: We compute the magnitude of azimuthal angular separation of two final state leptons as:
    \[\Delta \phi_{l_1 l_2}=|\phi_{l_1}-\phi_{l_2}| \text{  for $\Delta \phi_{l_1 l_2}<\pi$}\]
    \[\Delta \phi_{l_1 l_2}=2\pi-|\phi_{l_1}-\phi_{l_2}|, \text{  otherwise}\]

The normalized distributions of $ \Delta \phi_{l_1 l_2}$ are presented in Fig.\ref{fig1}(f)~(Fig.\ref{fig4}(e)). We choose  $\Delta \phi_{l_1 l_2}>3.0 (3.1)$, for the survival of the signal over background.

\item $C_7:$ We define the azimuthal angular separation between direction of the leading lepton $l_1$ and missing energy vector $\cancel{E}_T$ as $\Delta \phi_{l_1,~ \cancel{E}_T}$ and show the normalized distributions of the same in Fig.\ref{fig4}(f). To differentiate the signal from the background, most optimal cut on the aforementioned variable is $\Delta \phi_{l_1,~ \cancel{E}_T}>3.1$.
\end{itemize}

We summarize the results obtained after each cut, {\em i.e.} $(B_1- B_7)$ for $\sqrt{s}=6$ TeV and $(C_1- C_7)$ for $\sqrt{s}=14$ TeV mentioned above in two tables. The cut-flow of the first three benchmarks along with the SM background for COM 6 TeV are presented in Table \ref{tab:BP1_6TEV_2LMET}. The number of signal and background events after applying each cut are quoted at integrated luminosity 100 fb$^{-1}$. It is evident from the table that the cuts are very effective to wipe out the background compare to the signal. At $\sqrt{s} = 6$ TeV  and $\mathcal{L} = 100$ fb$^{-1}$, we are left with only 192 background events. Whereas the signal events remains almost steady after applying the same cuts. In the last column of the table, we compute the required integrated luminosity to achieve the 5$\sigma$ significance. For BP1, 0.2 ${\rm fb}^{-1}$ luminosity is enough to achieve $5\sigma$ significance while a benchmark with higher heavy neutrino mass 5.11 TeV (BP3) would require 9.1 ${\rm fb}^{-1}$. This happens owing to the decrease in signal cross section with increasing $M_{\tilde{\Psi}_{1}}$.


Next in Table \ref{tab:BP2_14TEV_2LMET}, we present the cut flow for benchmarks BP4-BP8 for COM energy 14 TeV with integrated luminosity 1600 ${\rm fb}^{-1}$. SM background events are reduced to 237 after employing all the chosen cuts. Whereas the number of the signal events surviving after imposing all cuts are 25809 (for BP4), 4698 (for BP5), 1323 (for BP6), 430 (for BP7), and 98 (for BP8). In addition, we derive the integrated luminosity needed to attain $5\sigma $ significance. For BP4-BP8, 2 ${\rm fb}^{-1}$, 9 ${\rm fb}^{-1}$, 36 ${\rm fb}^{-1}$, 144 ${\rm fb}^{-1}$ and 1381 ${\rm fb}^{-1}$ integrated luminosities are required to achieve 5$\sigma$ discovery. Thus one can conclude that the benchmark with larger cross section and hence with smaller value of $M_{\tilde{\Psi}_{1}}$, is more promising to probe at the muon collider than others.


\begin{table}[htpb!]
	\centering
		\resizebox{12cm}{!}{
	\begin{tabular}{|p{4cm}|c|c|c|c|c|c|c|p{4cm}|}
		\cline{2-8}
		\multicolumn{1}{c|}{}& \multicolumn{7}{|c|}{Number of Events after cuts ($\mathcal{L}=100$ fb$^{-1}$)} &  \multicolumn{1}{c}{} \\ \cline{1-8}
		 \multicolumn{1}{|c|} {SM-background } 
		 & $B_1$  &  $ B_2 $    &  $B_3$ & $B_4$   & $B_5$  & $B_6$  & $B_7$    & \multicolumn{1}{c}{}
		\\ \cline{1-8} 
           \multicolumn{1}{|c|} {$2l+\mET$} & 19040  & 4839 & 2230 & 1622 & 624 & 396 & 192   \\ \cline{1-7}
 
		\cline{1-7}  \hline
		                      
			\multicolumn{1}{|c|}{Signal }  &\multicolumn{7}{|c|}{} &  \multicolumn{1}{|c|}{$\mathcal{L}_{5\sigma}$ (fb$^{-1}$)} \\ \cline{1-9} 
		\multicolumn{1}{|c|}{BP1 } & 19176  & 18239  & 17427 & 16672 & 15348 & 15113 & 14656 &  \multicolumn{1}{|c|}{0.2}   \\ \hline
		\multicolumn{1}{|c|}{BP2 } & 2825   & 2683  & 2566  & 2440 & 2227 & 2195 & 2128 & \multicolumn{1}{|c|}{1.2} \\ \hline 
		\multicolumn{1}{|c|}{BP3 } & 393   & 373  &  355 & 336 & 306 & 302 & 292 & \multicolumn{1}{|c|}{9.1} \\ \hline 
		
	\end{tabular}}

	\caption{ The cut-flow for signal and backgrounds along with the significances for BP1, BP2 and BP3 at 6 TeV Muon collider and the required integrated luminosity for 5$\sigma$ significance for the $ \mu^+ \mu^- \rightarrow 2l+\mET$ channel. }
	\label{tab:BP1_6TEV_2LMET}
\end{table}	
\begin{table}[htpb!]
	\centering
		\resizebox{12cm}{!}{
	\begin{tabular}{|p{3.0cm}|c|c|c|c|c|c|c|p{3.0cm}|}
		\cline{2-8}
		\multicolumn{1}{c|}{}& \multicolumn{7}{|c|}{Number of Events after cuts ($\mathcal{L}=1600$ fb$^{-1}$)} &  \multicolumn{1}{c}{} \\ \cline{1-8}
		\multicolumn{1}{|c|} {SM-background } 
		 & $C_1$  &  $ C_2 $    &  $C_3$ & $C_4$   & $C_5$  & $C_6$  & $C_7$  &  \multicolumn{1}{c}{}
		\\ \cline{1-8} 
            \multicolumn{1}{|c|} {$2l+\mET$} & 239257  & 73455 & 13677 & 9070 & 5676 & 1094 & 237   \\ \cline{1-9}
 
		\cline{1-7}  \hline
\multicolumn{1}{|c|}{Signal }  &\multicolumn{7}{|c|}{} &  \multicolumn{1}{|c|}{$\mathcal{L}_{5\sigma}$ (fb$^{-1}$)} \\ \cline{1-9}		               
		                      
		\multicolumn{1}{|c|}{BP4 } & 54794   & 54637  & 53629 &  48035 & 47547 & 40009 & 25809 &  \multicolumn{1}{|c|}{2} \\ \hline 
		\multicolumn{1}{|c|}{BP5 } & 10079  & 10046 &  9878 & 8804 & 8709 & 7332 & 4698 &  \multicolumn{1}{|c|}{9}   \\ \hline
		\multicolumn{1}{|c|}{BP6 } & 2834  & 2825 & 2770  & 2460 & 2432 & 2052 & 1323 & \multicolumn{1}{|c|}{36}   \\ \hline
		\multicolumn{1}{|c|}{BP7 } & 947  & 944 & 925  & 810 & 801 & 672 & 430 &  \multicolumn{1}{|c|}{144}   \\ \hline
		\multicolumn{1}{|c|}{BP8 } & 217  & 216 & 211  & 186 & 184 & 154 & 98 &  \multicolumn{1}{|c|}{1381}   \\ \hline	
	\end{tabular}}

	\caption{ The cut-flow for signal and backgrounds along with the significances for BP4, BP5, BP6, BP7 and BP8 at 14  TeV Muon collider and the required integrated luminosity for 5$\sigma$ significance for the $ \mu^+ \mu^- \rightarrow 2l+\mET$ channel.}
	\label{tab:BP2_14TEV_2LMET}
\end{table}

\section{Summary and conclusion}
\label{conclusion}

In this present paper, we have focused on a framework popularly known as ISS(2,2), i.e. (2,2) inverse see-saw realisation, where the SM is accompanied with two singlet right handed neutrinos and two singlet neutral fermions. From the name of the model itself, it is evident that the neutrino mass generation in this scenario occurs via inverse see-saw mechanism. To make the analysis a robust one, we have considered the most general structure of the $7 \times 7$ neutrino mass matrix $M_\nu$ with complex entries. Further by regulating the input model parameters and assuming normal mass hierarchy between the active neutrinos, we can segregate the portion of the parameter space, which becomes compatible with the neutrino oscillation data. Upon diagonalising the mass matrix $M_\nu$, we are left with three active neutrino and four heavy neutrino mass eigenstates ($\tilde{\Psi_1}, \tilde{\Psi_2}, \tilde{\Psi_3}, \tilde{\Psi_4}$). The interesting feature of these heavy neutrino mass eigenstates is the pairwise mass degeneracy of the same, {\em i.e.}  $M_{\tilde{\Psi_1}} \simeq M_{\tilde{\Psi_2}}$ and $M_{\tilde{\Psi_3}} \simeq M_{\tilde{\Psi_4}}$. In addition, we have checked that the former parameter space also obeys the bounds coming from the LFV decays. Next at each and every point in the parameter space satisfying neutrino oscillation data and the bounds coming from the LFV decays, the baryon asymmetry is computed. Thus we can obtain a subset of the larger parameter space, which is compatible with (i) neutrino oscillation data, (ii) constraints coming from LFV decay and (iii) observed baryon asymmetry data. One can conclude that the baryon asymmetry constraint itself has put a lower bound on the lightest heavy neutrino state {\em i.e.} $M_{\tilde{\Psi_1}} > 3.2$ TeV.

Next we aim to analyse some promising collider signatures at multi-TeV muon collider. First we choose some benchmark points from the previously derived parameter space, depending on $M_{\tilde{\Psi_1}}$. For collider study, we intend to analyse those benchmarks which satisfy the observed baryon asymmetry of the universe. Accordingly, we have picked up different benchmarks where $M_{\tilde{\Psi_1}}$ ranges between 3.2 TeV to 10.3 TeV (BP1-BP8).The range of heavy neutrino masses which we are talking about, cannot be produced at ILC. The signal contains pair production of active neutrino and a heavy neutrino, followed by the decay of the heavy neutrino into $W^\pm$ and a charged lepton (e, $\mu$). Further if we consider the leptonic decay of the generated $W^\pm$, it will lead to a di-lepton plus missing energy state. For analysis, we choose $\sqrt{s} = 6$ TeV for BP1, BP2 and BP3 and $\sqrt{s} = 14$ TeV for the rest of the benchmarks (BP4-BP8). We have not considered conventional LHC or ILC for purpose. The cross-section of the same process at LHC is much smaller than what we obtained in muon-collider. In addition, the large LHC backgrounds would be sufficient to kill the signal. The disadvantage of ILC lies in the maximal achievable COM energy (1 TeV). The dominant background for this signal is $2l + \mET$, arising from the subprocesses like $W^+ W^-, ~ZZ, ~ W^+ W^- Z, ~ ZZZ$. Here we impose cuts on a few relevant kinematic variables to achieve maximally enhanced signal significances, suppressing the backgrounds with respect to the signals. Among all benchmarks, BP1 (BP4) turns out to be most promising with respect to the analysed signal at $\sqrt{s} =6$ TeV (14 TeV). To achieve 5$\sigma$ significance one needs 0.2 fb$^{-1}$, 1.2 fb$^{-1}$ and 9.1 fb$^{-1}$ integrated luminosity for BP1, BP2, BP3v respectively at COM energy 6 TeV. Besides the integrated luminosities required for 5$\sigma$ significance are 2 fb$^{-1}$, 9 fb$^{-1}$, 36 fb$^{-1}$, 144 fb$^{-1}$, 1381 fb$^{-1}$ respectively for BP4, BP5, BP6, BP7, BP8 at $\sqrt{s}=14$ TeV. Though this particular channel provides encouraging results at muon collider, one can also consider the hadronic decay of $W^\pm$ in the signal instead of leptonic decay, leading to $1l + 2j + \mET$ signature. Analysing this channel could be challenging due to the presence of large jetty SM backgrounds. We shall present a comprehensive analysis of this channel separately for comparison.

\section{ACKNOWLEDGEMENTS}

The authors thank Aleksander Filip Zarnecki, Yongcheng Wu,  Nabarun Chakrabarty and Gourab Saha for technical help and fruitful discussions. IC acknowledges support
from DST, India, under grant number IFA18-PH214 (INSPIRE Faculty Award). HR is supported by the Science and Engineering Research Board, Government of India, under
the agreement SERB/PHY/2016348 (Early Career Research Award). TS acknowledges the support from the Dr. D. S. Kothari Postdoctoral scheme No. PH/20-21/0163. 

\appendix
\section{Evolution of asymmetry}
\label{app:A} Evolution of asymmetry can be calculated by solving the following Boltzmann equations:
 $\tilde{\Psi_{1}}, \tilde{\Psi_{2}}$ and  the $(B-L)$ asymmetry can be written as \cite{Plumacher:1996kc},
 \bea
 \frac{\text{d} Y_{\tilde{\Psi_{1}}}}{\text{d} z} &=& - \frac{z}{s \hspace{1mm} H(M_{\tilde{\Psi_1}})} \Big[ \Big(\frac{Y_{\tilde{\Psi_{1}}}}{Y^{eq}_{\tilde{\Psi_{1}}}} - 1\Big)(\gamma_{D}^{(1)} + 2 \gamma^{(1)}_{\phi,s} + 4 \gamma^{(1)}_{\phi,t})\Big] \,,
  \label{boltzmann-eq1}
 \eea
  \bea
 \frac{\text{d} Y_{\tilde{\Psi_{2}}}}{\text{d} z} &=& - \frac{z}{s \hspace{1mm} H(M_{\tilde{\Psi_1}})} \Big[ \Big(\frac{Y_{\tilde{\Psi_{2}}}}{Y^{eq}_{\tilde{\Psi_{2}}}} - 1\Big)(\gamma_{D}^{(2)} + 2 \gamma^{(2)}_{\phi,s} + 4 \gamma^{(2)}_{\phi,t})\Big] \,,
  \label{boltzmann-eq2}
 \eea
  \bea
 \frac{\text{d} Y_{B-L}}{\text{d} z}  &=& - \frac{z}{s \hspace{1mm} H(M_{\tilde{\Psi_1}})} \Big[ \sum_{j=1}^2 \left\{\frac{1}{2} \frac{Y_{B-L}}{Y^{eq}_{l}} + \epsilon_j ~\Big(\frac{Y_{\tilde{\Psi_{j}}}}{Y^{eq}_{\tilde{\Psi_{j}}}} - 1\Big) \right\} \gamma_{D}^{(j)}  \nonumber \\
 && +\frac{Y_{B-L}}{Y^{eq}_{l}}\left\{2 \gamma_{\tilde{\Psi},s} + 2 \gamma_{\tilde{\Psi},t}\right\}+ \frac{Y_{B-L}}{Y^{eq}_{l}} \sum_{j=1}^2 \left\{ 2 \gamma^{(j)}_{\phi,t} + \frac{Y_{\tilde{\Psi_{j}}}}{Y^{eq}_{\tilde{\Psi_{j}}}}  \gamma^{(j)}_{\phi,s} \right\} \Big] \,,
 \label{boltzmann-eq3}
 \eea
where $z = \frac{M_{\tilde{\Psi_1}}}{T}$ and $H(M_{\tilde{\Psi_1}})$ is the Hubble parameter at $T = M_{\tilde{\Psi_1}}$ and $H(T=M_{\tilde{\Psi_1}}) = 1.66 ~g_{eff}^{1/2} \frac{T^{2}}{M_{\rm{Pl}}}|_{T=M_{\tilde{\Psi_1}}} $, $M_{\rm{Pl}} \sim 10^{19}$ GeV being Planck scale. $Y_{\tilde{\Psi_{j}}}^{eq}, Y_l^{eq}$ are the comoving densities at equilibrium. We solve these three equations with initial conditions :
\bea
Y_{\tilde{\Psi_i}}(0) = Y_{\tilde{\Psi_i}}^{eq} , ~{\rm and}~ Y_{B-L}(0) = 0 \,.
\eea
at $T >> M_{\tilde{\Psi_1}}$.

\begin{figure}[htpb!]{\centering
\subfigure[]{
\includegraphics[scale=0.1]{pshi1_l_phi.png}}
\subfigure[]{
\includegraphics[scale=0.1]{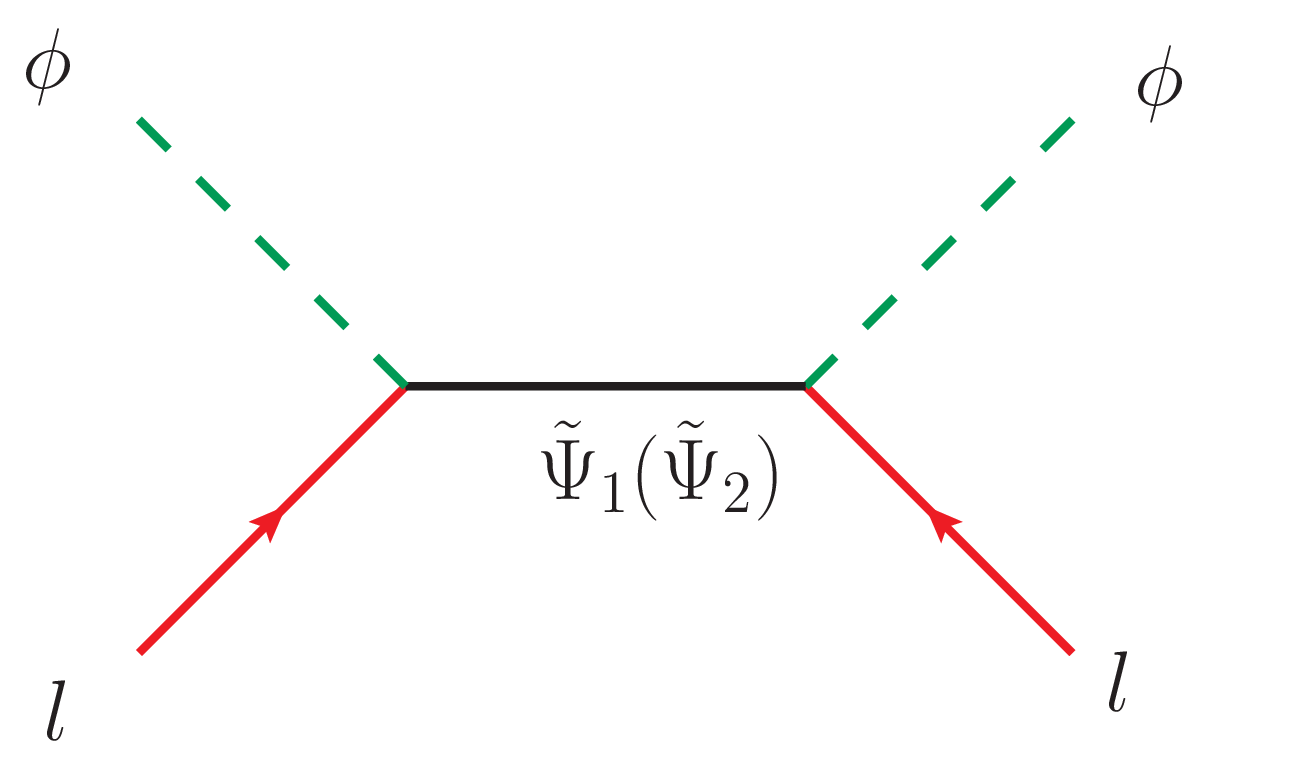}}
\subfigure[]{
\includegraphics[scale=0.1]{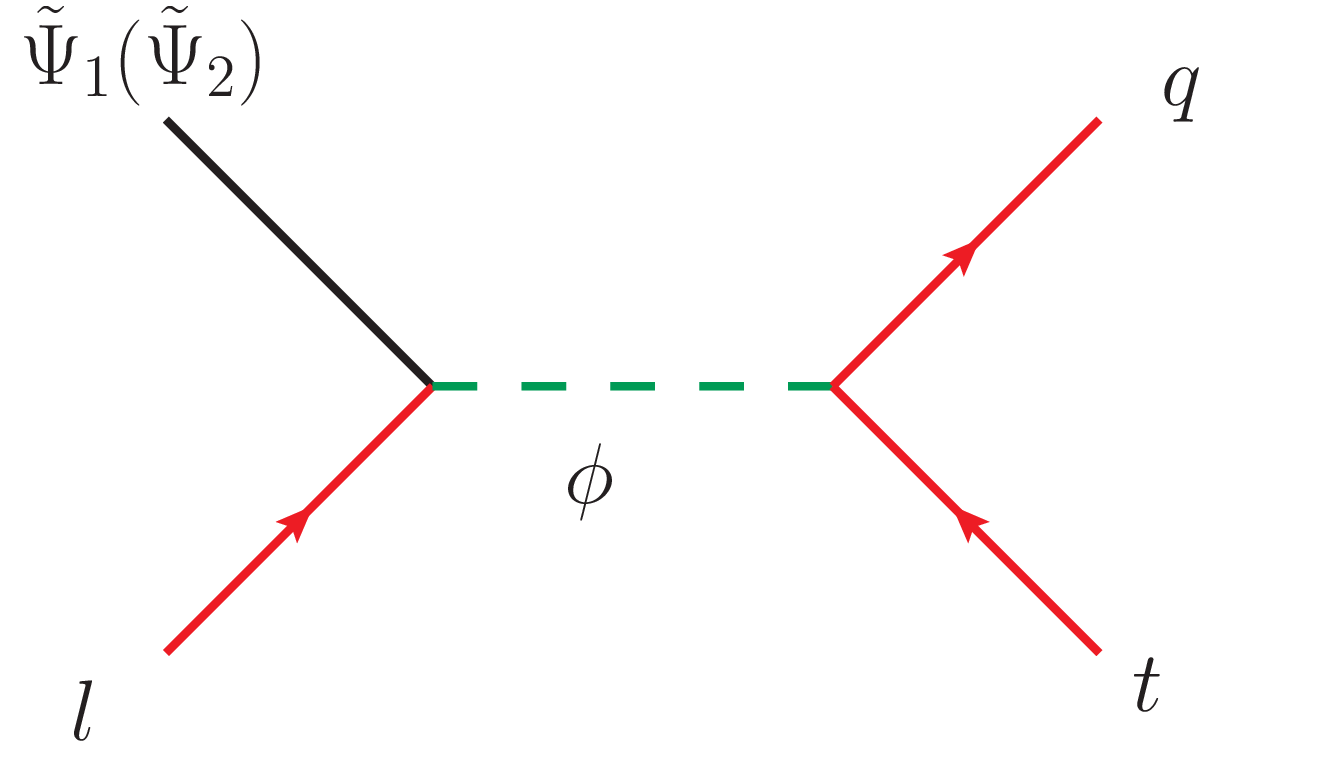}} \\
\subfigure[]{
\includegraphics[scale=0.1]{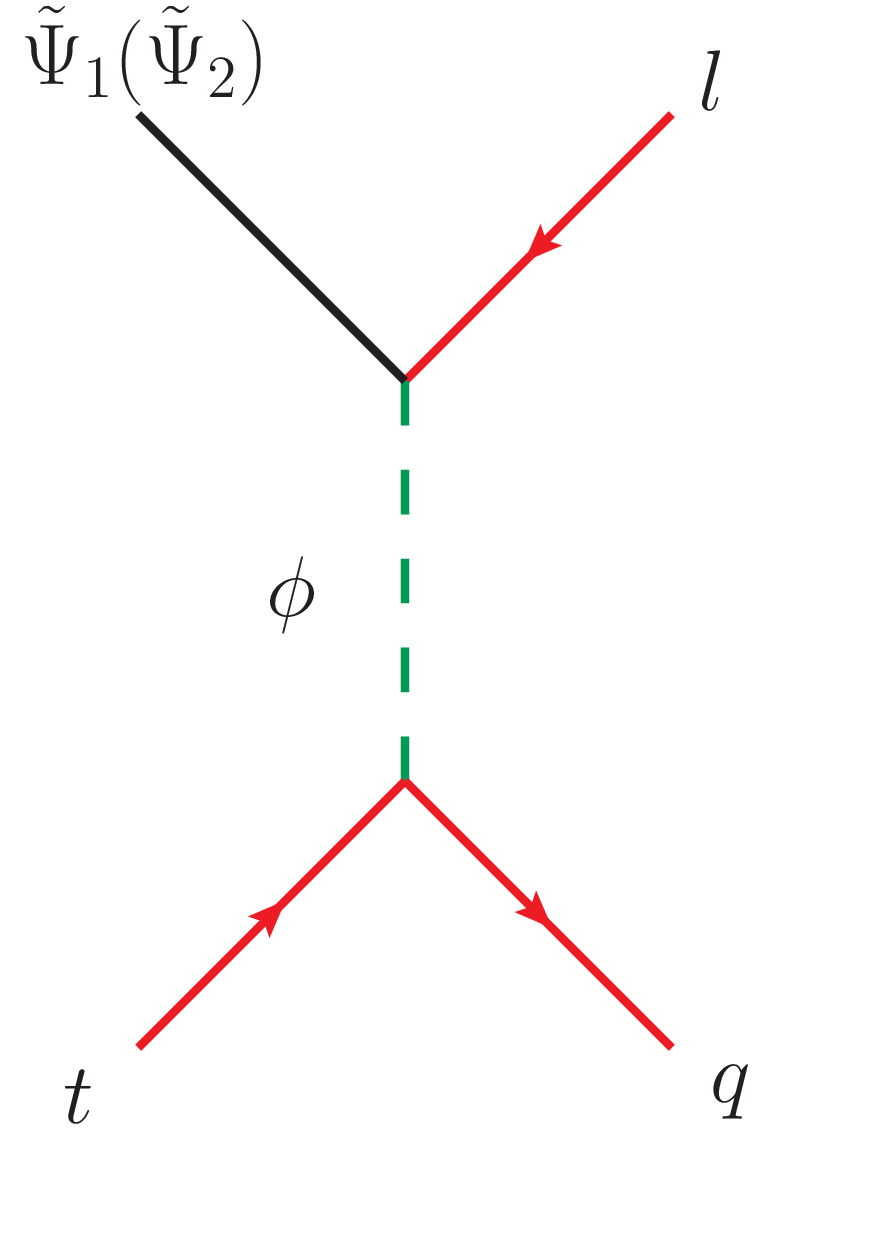}}
\subfigure[]{
\includegraphics[scale=0.1]{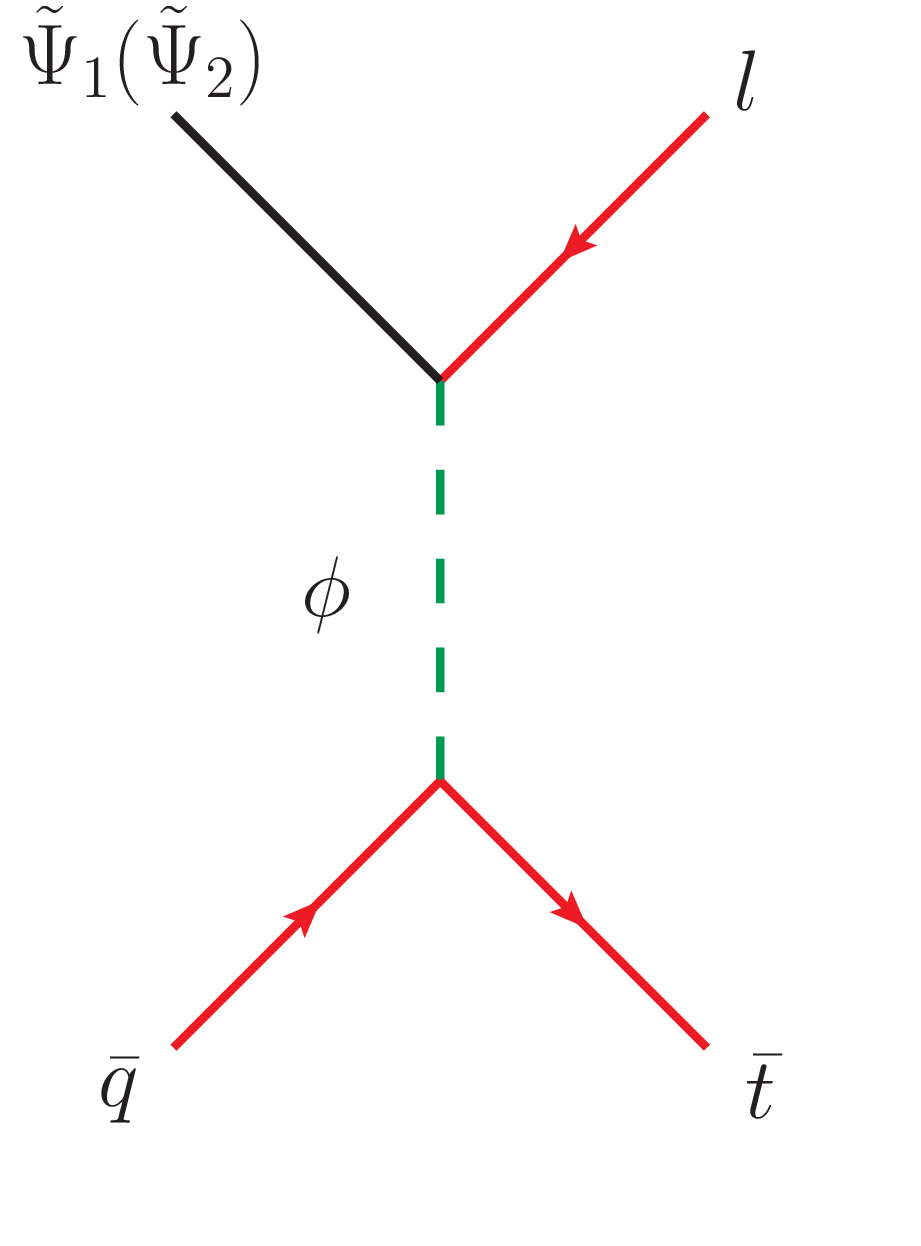}}
\subfigure[]{
\includegraphics[scale=0.1]{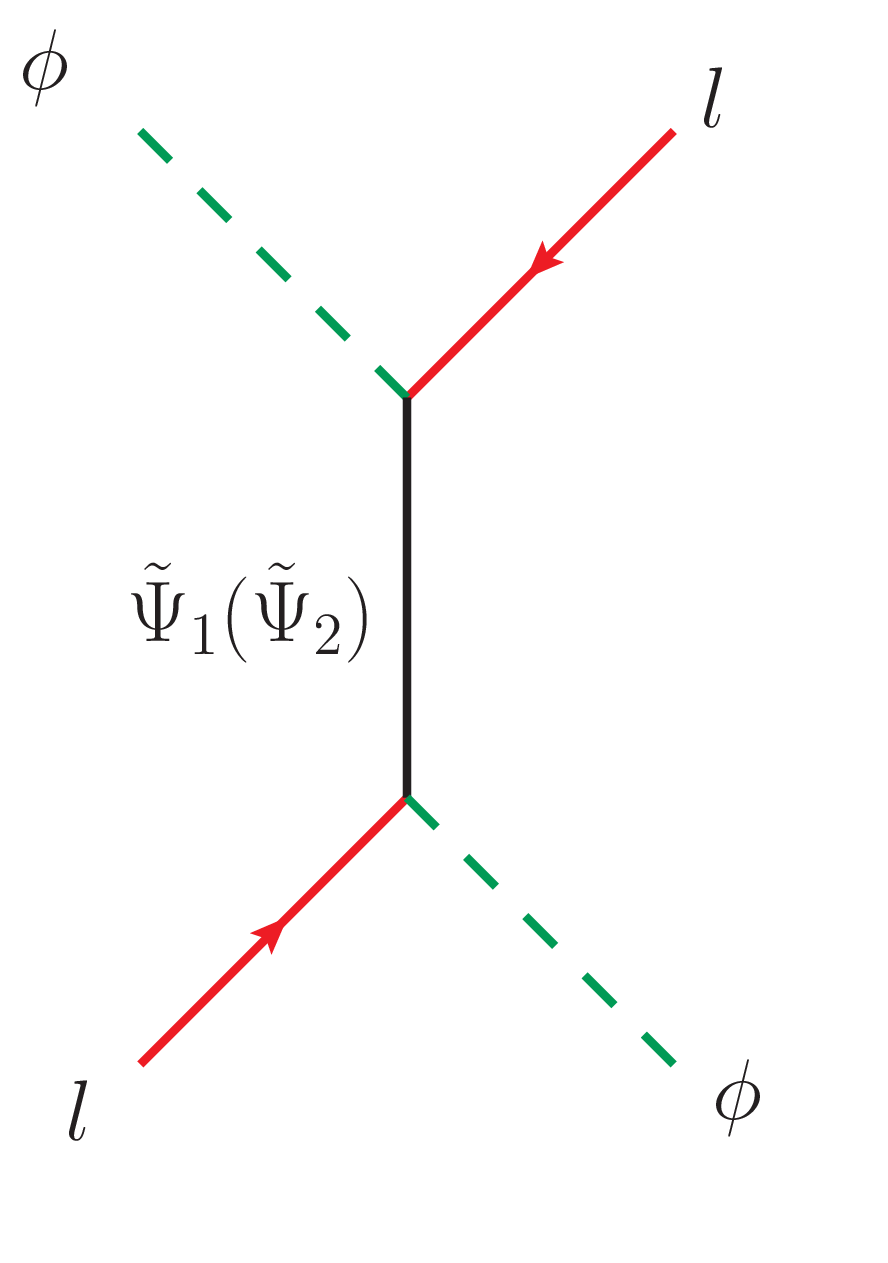}}} 
\caption{Feynman diagram representing  (a) decay of lightest heavy RH-neutrino $\tilde{\Psi_{1}}~(\tilde{\Psi_{2}})$ (contributes to $\gamma_{D}^{(1)}, \gamma_{D}^{(2)}$), (b) $\Delta L = 2$, $s$-channel scattering via $\tilde{\Psi_{1}}~(\tilde{\Psi_{2}})$ (contributes to $\gamma_{\tilde{\Psi},s}$), (c) $\Delta L = 1$, $s$-channel scattering via Higgs (contributes to $\gamma^{1}_{\phi,s}, \gamma^{2}_{\phi,s}$), (d) and (e) $\Delta L = 1$, $t$-channel scattering via Higgs (contributes to $\gamma^{1}_{\phi,t}, \gamma^{2}_{\phi,t}$), (f) $\Delta L = 2$, $t$-channel scattering via $\tilde{\Psi_{1}}~(\tilde{\Psi_{2}})$ (contributes to $\gamma_{\tilde{\Psi},t}$). }
\label{diagram}
\end{figure}
Contributions to $\gamma^{'}$s are given by diagrams shown in Fig.~\ref{diagram}.


The lepton asymmetry is further can be translated into baryon asymmetry.  This results in the final baryon number at $T_{\rm sph} \sim 150$ GeV (the freeze-out temperature of the sphelaron process ) as \cite{Burnier:2005hp}:
\bea
Y_{B} = \bigg( \frac{8 N_{f} + 4 N_{H}}{22 N_{f} + 13 N_{H} } \bigg) Y_{B-L} (z_{\rm sph}).
\label{baryon-asym-con}
\eea
With $Y_{B-L} (z_{\rm sph})$ as the solution of Boltzmann equations at $z = z_{\rm sph} = \frac{M_{\tilde{\Psi_1}}}{T_{\rm sph}}$.

Here $N_f:$ the number of generations of fermion families
and $N_H:$ number of Higgs doublets. For our case, $N_f = 3$ and $N_H =1$.
$\gamma_{D}^{(i)}$ can be written as \cite{Plumacher:1996kc} :
\bea
\gamma_{D}^{(i)} = N_{\tilde{\Psi}_i}^{eq}~ \frac{K_{1}(\frac{M_{\tilde{\Psi}_i}}{T})}{K_{2}(\frac{M_{\tilde{\Psi}_i}}{T})}~ \Gamma_{\tilde{\Psi_i}} \,, {\rm with}~ i=2 
\eea
$N_{\tilde{\Psi}_i}^{eq}$ being the equilibrium number density of mass eigenstate $\tilde{\Psi}_i$. Here $K_1$ and $K_2$ are the first and second modified Bessel functions of second kind respectively and $\Gamma_{\tilde{\Psi}_i}$ is the total decay width of $\tilde{\Psi}_i$.

Decay width of $\tilde{\Psi}_i$ at tree level, 
\bea
\Gamma_{\tilde{\Psi}_i} &:= & \Gamma (\tilde{\Psi}_i \rightarrow \phi^{\dagger} + l) + \Gamma (\tilde{\Psi}_i \rightarrow \phi + \bar{l}) \nonumber \\
&& = \frac{\alpha}{\text{sin}^{2}\theta_W} \frac{M_{\tilde{\Psi}_i}}{4} \frac{(M^{\dagger}_{D} M_{D})_{ii}}{M^{2}_{W}}
\label{decal-width}
\eea
with $\alpha, \theta_W$ being the Fine structure constant and the Weinberg angle.

For two body scattering $ a + b \rightarrow i + j + ...$, $\gamma_{eq}$ can be written as \cite{Plumacher:1996kc}, 
\bea
\gamma_{eq} = \frac{T}{64 \pi^{4}} \int_{(M_a+M_b)^2}^{\infty} \hspace{1mm} ds ~\hat{\sigma}(s) \hspace{1mm} \sqrt{s} \hspace{1mm}  K_{1}(\frac{\sqrt{s}}{T}) \,.
\eea
here $s$ \footnote{Not to be confused with "$s$-channel" mentioned earlier.} is the square of center of mass energy and $\hat{\sigma}(s)$ is reduced cross section, which can be expressed in terms of actual cross section as \cite{Plumacher:1996kc} :
\bea
\hat{\sigma}(s) = \frac{8}{s}\left[(p_a.p_b)^2 - M_a^2 M_b^2\right] \sigma(s) \,,
\eea 
with $p_k$ and $M_k$ being three momentum and mass of particle $k$.
Expressions for reduced cross-section ($\hat{\sigma}$) for all the scattering processes shown in Fig.~\ref{diagram} are given in \cite{Plumacher:1996kc,Chakraborty:2021azg}.

For our model the total D.O.F turns out to be :
\bea
g_{eff}~( T > 174 ~{\rm GeV}) &=& g_{\rm boson}+g_{\rm fermion} \nonumber \\
&&28 + \frac{7}{8} \times 94 
= 110.25
\eea

\bibliographystyle{JHEP}
\bibliography{refer1.bib} 
\end{document}